\RequirePackage{fix-cm}
\documentclass[twocolumn]{svjour3}  

\def\makeheadbox{\relax}

\makeatletter
\def\@seccntformat#1{\@ifundefined{#1@cntformat}%
  {\csname the#1\endcsname\quad}  
  {\csname #1@cntformat\endcsname}
}
\let\oldappendix\appendix 
\renewcommand\appendix{%
    \oldappendix
    \newcommand{\section@cntformat}{\appendixname~\thesection\quad}
}
\makeatother

\smartqed  

\usepackage{bbm,setspace,amssymb,amsfonts,mathrsfs,amsmath,mathtools,stmaryrd,bm,bm,etoolbox,comment,hyperref,url,color,enumitem,algorithm,booktabs,microtype,nicefrac,lingmacros,nccmath,tree-dvips,tikz-cd,lineno,physics,soul}

\usepackage{graphicx}
\usepackage{subcaption}
\usepackage{appendix}

\DeclareMathOperator*{\argmax}{argmax}
\DeclareMathOperator*{\loggrad}{log-grad}
\DeclareMathOperator{\E}{\mathbb{E}}

\newcommand{\bs}[1]{\bm{#1}} 

 \journalname{Statistics and Computing}
\begin{document}

\title{Optimally adaptive Bayesian spectral density estimation for stationary and nonstationary processes}


\author{Nick James         \and
        Max Menzies 
}


\institute{N. James \at
              School of Mathematics and Statistics \\
              The University of Melbourne\\
              VIC, 3010, Australia\\
              \email{nick.james@unimelb.edu.au}           
           \and
           M. Menzies \at
              Beijing Institute of Mathematical Sciences and Applications \\
              Tsinghua University\\
              Beijing, 101408, China\\
              \email{max.menzies@alumni.harvard.edu}           
}

\date{Received: date / Accepted: date}

\maketitle

\begin{abstract}
This article improves on existing methods to estimate the spectral density of stationary and nonstationary time series assuming a Gaussian process prior. By optimising an appropriate eigendecomposition using a smoothing spline covariance structure, our method more appropriately models data with both simple and complex periodic structure. We further justify the utility of this optimal eigendecomposition by investigating the performance of alternative covariance functions other than smoothing splines. We show that the optimal eigendecomposition provides a material improvement, while the other covariance functions under examination do not, all performing comparatively well as the smoothing spline. During our computational investigation, we introduce new validation metrics for the spectral density estimate, inspired from the physical sciences. We validate our models in an extensive simulation study and demonstrate superior performance with real data.


\end{abstract}




\section{Introduction}

Spectral density estimation (SDE) is a common method to understand the autocovariance structure of a stationary time series and perhaps the key technique to detect periodicities in time series data. Of particular importance in the physical sciences, such as mass spectrometry \cite{ToddChem,ToddChem2}, are the peaks of a power spectrum, and the frequencies at which they occur. However, most real-world processes are nonstationary. This has necessitated the development of methods for SDE for nonstationary time series. In this article, we improve upon existing Bayesian techniques to estimate time-varying power spectra.

Our analysis takes place in a nonparametric Bayesian framework and implements a \emph{reversible jump Markov chain Monte Carlo} (RJMCMC) algorithm, assuming a \emph{Gaussian process} (GP) prior. Aligning with the existing literature on locally stationary time series \cite{Dahlhaus1997,Adak1998}, we partition a given time series into a finite number of stationary segments. The true log power spectrum is estimated within the Whittle likelihood framework \cite{Whittle1957} and appropriate optimisation, using the local log periodograms. In doing so, no parametric assumption is made about the time series data itself (time domain), only about the log periodograms (frequency domain).

Our key improvement over existing techniques is the optimisation of an appropriate eigendecomposition, where a smoothing spline covariance structure is used. Therein, we maximise the Whittle likelihood with respect to the number of eigenvalues. This produces a suitable estimate of the power spectrum for both simple and complex spectra. To demonstrate the robustness of this smoothing spline optimisation, we also investigate alternative GP covariance functions. We specify to the stationary case and validate our estimates against analytically known autoregressive (AR) processes. We show that other covariance functions provide no benefit over the eigendecomposition of the smoothing spline. In the process thereof, we introduce new validation metrics that more appropriately reflect the needs of the physical sciences, where the amplitude and frequency of the peaks is paramount.

Our article builds off a rich history of nonparametric frequentist and Bayesian modelling to analyse and fit data. First, smoothing splines are the bedrock for such analysis. Cogburn and Davis \cite{Cogburn1974} and Wahba \cite{Wahba1980} pioneered the use of smoothing splines for spectral density estimation in the stationary case, with improvements made by Gandgopadhyay et al. \cite{Ganngopadhyay1999}, Choudhuri et al. \cite{Choudhuri2004} and others \cite{Wood2002}. Guo et al. \cite{Guo2003} used smoothing spline ANOVA in the locally stationary case. Variations of smoothing splines, such as B-splines \cite{Eilers1996}, P-splines \cite{Wood2017} and others \cite{Wahba1990,Gu2002}, have been used across a range of implementations in the statistics literature.

Alternative covariance functions for GPs, other than smoothing splines, have been used in a wide range of applications to model and analyse data. Rasmussen and Williams \cite{Rasmussen2006} provide an overview of modelling complex phenomena with combinations of covariance functions in GPs. Paciorek and Schervish \cite{Paciorek2004} perform GP regression and investigate smoothness over a class of nonstationary covariance functions, while Plagemann et al. \cite{Plagemann2008} utilise such functions within a Bayesian regression framework. More recently, nonparametric regression methods with automatic kernel learning have produced a GP regression framework for kernel learning \cite{Duvenaud2013}, and other frameworks \cite{Lu2016,Wilson2013}. However, alternative covariance functions have been used infrequently for the problem of SDE.

More recently, smoothing splines and GPs have been used to study the spectra of nonstationary processes. Whereas the work of Wahba \cite{Wahba1980} and Carter and Kohn \cite{Kohn1997} places priors over spectra of stationary processes, more recent work has extended this to nonstationary time series, predicated on the work of Dahlhaus \cite{Dahlhaus1997} and Adak \cite{Adak1998}. Reversible jump MCMC algorithms have been used to partition nonstationary time series into stationary segments. The number and location of these can be estimated flexibly from the data, and the uncertainty quantified, both in the time \cite{Wood2011,james2021_MJW,arjun} and frequency domain \cite{Rosen2009,Rosen2012}.

In particular, Rosen, Wood and Stoffer have used a smoothing spline GP covariance structure and RJMCMC algorithms in conjunction effectively to perform SDE in the nonstationary setting. Their work partitions nonstationary time segments into locally stationary segments, first with predetermined partitions \cite{Rosen2009} and then adaptively \cite{Rosen2012}, in each case using a product of local Whittle likelihoods to estimate the time series' time-varying power spectrum. The model includes a particular eigendecomposition of the covariance matrix associated to the smoothing spline. For computational savings, the first 10 eigenvectors are used - this number is fixed.

The main limitation of the aforementioned work is the global degree of smoothness induced by the fixed eigendecomposition. In our article's experiments, we show that this model generally underfits real-world processes. These usually exhibit more complexity than the autoregressive processes that are canonically used in the validation of such SDE models. This leads to a systemic bias in the literature towards models that produce excessively smooth spectral estimates. Indeed, Hadj-Amar et al. \cite{HadjAmar2019} examined the 10-eigenvector model of Rosen et al. \cite{Rosen2012} and concluded that the fixed decomposition tends to produce estimates that are overly smooth. Our article ameliorates this problem by varying the number of eigenvalues used and optimising it by maximising the Whittle likelihood. Quantitatively, we produce better results than other models as measured by new and existing validation metrics, and qualitatively, our estimates provide a more appropriate level of smoothness upon visual inspection.

The article is structured as follows: in Section \ref{sec:model and priors}, we introduce the mathematical model, including our key improvement. In Section \ref{sec:validationmetrics}, we describe the validation metrics we use in the paper to evaluate our spectral density estimates, including existing metrics and a new framework for measuring distances between non-trivial sets of peaks. In Section \ref{simulation}, we describe our simulation study, incorporating both our chosen alternative covariance functions and introducing our new validation metrics. In Section \ref{real data}, we apply our method to real data, and observe more appropriate fitting of our model to real-world data with complex periodograms. In Section \ref{conclusion}, we summarise our findings regarding the performance and appropriate smoothness produced by eigendecomposition optimisation versus variation of covariance functions in our GP model.

\section{Mathematical model}
\label{sec:model and priors}
\subsection{Spectral density estimation of stationary time series}
Let $(X_{t})_{t=1,...,n}$ be a discrete real valued time series. 
\begin{definition}
$(X_t)$ is stationary if
\begin{enumerate}
    \item Each random variable $X_t$ is integrable with a finite common mean $\mu$ for each $t$. By subtracting the mean, we may assume henceforth $\mu=0$.
    \item The autocovariance  $\E[(X_{t}-\mu)(X_{t+k}-\mu)]$ is a function only of $k$, which we denote $\gamma(k)$.
\end{enumerate}
\end{definition}
We shall reserve the letter $n$ to denote the length of a stationary time series, and $T$ to denote the length of a not-necessarily-stationary time series. Spectral analysis allows us to study the second-order properties, particularly periodicity, of a stationary time series expressed in the autocovariance structure. The power spectral density function (PSD) is defined as follows: 
 \begin{equation}
     f(\nu) = \sum^{\infty}_{k=-\infty} \gamma (k) \exp (-2\pi i\nu k), \text{ for} -\frac{1}{2} \leq \nu \leq \frac{1}{2}.
 \end{equation}
Most important are the \emph{Fourier frequencies} $\nu_j=\frac{j}{n}, j=0,1,...,n-1$. Define the \emph{discrete Fourier transform} (DFT) of the time series:
\begin{equation}
Z(\nu_j) =\frac{1}{\sqrt{n}} \sum_{t=1}^n X_{t} \exp(-2 \pi i \nu_j t), \text{ for } j = 0, ... ,n-1.
\end{equation}
Whittle initially showed that under certain conditions, the DFTs have a \emph{complex normal} (CN) distribution \cite{Whittle1957,Whittle1954}: 
\begin{equation}
 \mathbf{Z}\sim \text{CN}(0, \mathbf{F}), \text{ where } \mathbf{F}=\text{diag}[f(\nu_{0}),...f(\nu_{n-1})].
\end{equation}
For each Fourier frequency component, a noisy but unbiased estimate of the PSD $f(\nu_{k})$ is the \emph{periodogram} \cite{Choudhuri2004}, defined by $I(\nu_{j})=|Z(\nu_j)|^2$. By symmetry, the periodogram contains $m=[\frac{n}{2}]+1$ effective observations, corresponding to  $j=0,1,...,m-1$. Rosen et al. \cite{Rosen2009,Rosen2012} outline a signal plus noise representation of the periodogram:
\begin{equation}
\label{eq:log_periodigram}
\log I(\nu_{j}) = \log f(\nu_{j}) + \epsilon_{j}, 
\end{equation}
where $\epsilon_{j}$ is a $\log(\text{Exp}(1))$ random variable. Recall that a random variable $Y$ is said to have distribution $\text{Exp}(1)$ if it has cumulative density function $\mathbb{P}(Y \leq y) = 1 - e^{-y}$. With this definition, the representation (\ref{eq:log_periodigram}) assumes each noise variable $\epsilon_{j}$ (noise in the log periodogram) is an independent and identically distributed (i.i.d.) whose distribution is identical to $\log(Y)$. This model is justified by the following result: for a wide class of theoretical models, Theorem 10.3.2 of \cite{Brockwell1991} proves that the vector of quotients 
\begin{equation}
\Big( \frac{I(\nu_0)}{f(\nu_0)} ,..., \frac{I(\nu_{m-1})}{f(\nu_{m-1})}  \Big)    
\end{equation}
converges in distribution to a vector of i.i.d. $\text{Exp}(1)$ random variables. We reiterate that this theorem applies to the noise variables in the log periodogram, not the original time series. In particular, we make use of no parametric assumptions regarding the underlying process or noise of the original time series (time domain), and only of the log periodogram (frequency domain) as specified by (\ref{eq:log_periodigram}).

For our applications, we assume the quotient $\frac{I}{f}$ is approximately exponentially distributed with mean $1$. This is a simpler representation of the second moment of the distribution, reducing the problem of covariance estimation to a simpler problem of mean estimation. Note the periodogram oscillates around the true spectral density, so there is a delicate balance between inferring the spectrum of a process and excessive smoothing, leading to negligible inference. We remark that unlike most statistical models, this does not assume Gaussian noise; in fact, by the aforementioned theoretical results, we assume exponentially distributed random noise instead.

\subsection{Gaussian process regression}
\begin{definition}
\label{def:GP}
A \emph{Gaussian process} is a collection of random variables, any finite collection of which have a multivariate Gaussian distribution \cite{Rasmussen2006}. Such a process $f(\mathbf{x})$ is uniquely determined by its mean function $m(\mathbf{x}) = \E[f(\mathbf{x})]$ and covariance function 
$k(\mathbf{x},\mathbf{x'}) = \E[(f(\mathbf{x})-m(\mathbf{x})) (f(\mathbf{x'})-m(\mathbf{x'})]$. We write $f(\textbf{x}) \sim \mathcal{GP}(m(\mathbf{x}),k(\mathbf{x},\mathbf{x'}))$.
\end{definition}
We assume the standard ``noisy observation'' Gaussian additive error model $y = f(\mathbf{x}) + \epsilon$, in which $\epsilon$ is i.i.d Gaussian noise with variance $\sigma_{n}^{2}$ \cite{Rasmussen2006}. Then, the prior over the noisy observations is $\text{cov}(\mathbf{y}) = k(\mathbf{x,x'}) + \sigma_{n}^{2}I$. The joint distribution of observed response values, $\mathbf{y}$,  and function values evaluated at test points under the prior distribution, $\mathbf{f_{*}}$ is expressed as 
    \begin{equation}\nonumber
      \left[
        \begin{array}{c}
          \mathbf{y} \\
          \mathbf{f_{*}}
        \end{array}
        \right] 
      \sim \mathcal{N}
      \left(
        \begin{array}{c}
        \mathbf{0},  
        \end{array}
      \left[
        \begin{array}{cc}
          k(\mathbf{x,x})+\sigma^2_nI & k(\mathbf{x,x_*}) \\
          k(\mathbf{x_*,x}) & k(\mathbf{x_*,x_*})
        \end{array}
        \right]
      \right).
    \end{equation}
Finally, the GP regression predictive equations are as follows \cite{Rasmussen2006}:
\begin{equation}
    \mathbf{f_{*}}|\mathbf{x},\mathbf{y}, \mathbf{x'} \sim \mathcal{N}(\bar{\mathbf{f_*}},\text{cov}(\mathbf{f_{*}})),
\end{equation}
where the mean and covariance are defined as
\begin{align}
    \bar{\mathbf{f_*}} &:= \E[\mathbf{f_{*}}|\mathbf{x},\mathbf{y}, \mathbf{x_*}] = k(\mathbf{x_*,x})[k(\mathbf{x,x}) + \sigma_{n}^{2} I ]^{-1}\mathbf{y}, \\
    \text{cov}(\mathbf{f_{*}}) &:= k(\mathbf{x_*,x_*}) - k(\mathbf{x_*,x})[k(\mathbf{x,x}) + \sigma_{n}^{2}]^{-1} k(\mathbf{x,x_*}).
\end{align}

\subsection{Model and priors: stationary case}
\label{sec:stationarymodel}

We follow \cite{Choudhuri2004,Rosen2012} and use the Whittle likelihood function to model the log periodogram within a Bayesian regression framework. Let $\mathbf{y} = \left(y(\nu_{0}),...,y(\nu_{m-1})\right)$ be the log of the periodogram, $ \mathbf{f}=\left(f(\nu_{0}),...,f(\nu_{m-1})\right)$ be the PSD, and $\mathbf{g}=\left(g(\nu_{0}),...,g(\nu_{m-1})\right)$ be the log of the PSD. The likelihood of the log periodogram given the true power spectrum  can be approximated by
\begin{equation}
\label{eq:stationarywhittle}
p(\mathbf{y}| \mathbf{f}) = (2 \pi)^{-m/2} \prod_{j=0}^{m-1} \exp\left({-\frac{1}{2}\left[\log f(\nu_{j}) + \frac{I(\nu_{j})}{f(\nu_{j})}\right]}\right).
\end{equation}
Following \cite{Rosen2009,Rosen2012} we rewrite Eq.  (\ref{eq:log_periodigram}) as
\begin{equation}
y(\nu_{j}) = g(\nu_{j}) + \epsilon_{j},
\end{equation}
To obtain a flexible estimate of the spectrum, we place a GP prior over the unknown function $\mathbf{g}$. That is, we assume $\mathbf{g} \sim N(\mathbf{0},k(\mathbf{x,x_*}))$. This is determined by its covariance function. Most of the SDE literature to date has used smoothing splines \cite{Wahba1980} or other spline varieties for this covariance structure. In subsequent sections, we investigate both smoothing splines and other covariance functions. We choose kernels that have been frequently examined elsewhere in the literature: the squared exponential, Matern$_{3/2}$, Matern$_{5/2}$, Sigmoid kernel, and combinations of these such as the squared exponential + Sigmoid  \cite{Rasmussen2006,Paciorek2004,Plagemann2008,Lu2016,Wilson2013}.

\subsection{Nonstationary model and priors}

In this section, we describe the model assumed throughout the paper and associated priors. For the majority of this paper, the precise form of nonstationarity we assume is a Dahlhaus piecewise stationary process \cite{Dahlhaus1997}, which can be described as follows. Let a time series of length $T$, $(X_t)_{t=1,...,T}$, consist of an unknown number of segments $m$ and change points $\tau_1,...,\tau_{m-1}$ between segments. For notational convenience, set $\tau_0=0, \tau_m=T$. Then the entire time series $X_t$ can be written
\begin{align}
    \label{eq:nonstationarymodel}
    X_t = \sum_{i=1}^m \bm{1}_{[\tau_{i-1}+1,\tau_i]} X_t^i,
\end{align}
where each $X_t^i$ is an independent and stationary time series over the interval $[\tau_{i-1}+1,\tau_i]$. Following the notation of Section \ref{sec:stationarymodel}, let $n_i = \tau_i - \tau_{i-1}$ be the length of the $i$th segment, $m_i  =[\frac{n_i}{2}]+1$ be the effective length of each periodogram, and let $\mathbf{f}_i \in \mathbb{R}^{m_i}$ denote the PSD of the stationary time series $X_t^i$.

By independence of the processes, the Whittle likelihood approximation of the nonstationary model for a given partition $\boldsymbol{\tau}=(\tau_1,...,\tau_{m-1})$ is as follows:
\begin{align}
    \label{eq:nonstationarywhittle}
    L(\mathbf{f}_1,...,\mathbf{f}_m | \boldsymbol{\tau}, \mathbf{y}_1,...,\mathbf{y}_m) = \prod_{i=1}^m p(\mathbf{y}_i| \mathbf{f}_i),
\end{align}
where each local Whittle likelihood $p(\mathbf{y}_i| \mathbf{f}_i)$ is calculated according to (\ref{eq:stationarywhittle}).

Following Rosen et al. \cite{Rosen2012}, we place the following prior distributions on both the number of segments $m$ and the segment partitions $\boldsymbol{\tau} \in \mathbb{R}^{m-1}$.

\begin{enumerate}
    \item The prior on the number of segments $m$ is an integer-valued uniform distribution with minimal number 1 and maximum number $M$.
    
    \item The prior on each change point $\tau_i, i>0$ is a uniform distribution conditional on the previous change point $\tau_{i-1}$. Specifically, we allow $\tau_i$ to take any value with equal probability within a range $[\tau_{i-1} + t_{\text{min}}, T]$. This requires the specification of a minimal possible segment length $t_{\text{min}}$, which we set to be 40.
    
    \item The priors on each log PSD $\mathbf{g}_i= \log \mathbf{f}_i, i=1,...,m$ are independent and are specified as in Section \ref{sec:stationarymodel}.
\end{enumerate}

\subsection{MCMC implementation and choice of covariance functions}

The primary contribution of this paper is an improved method for SDE of nonstationary time series. Our work uses the RJMCMC of Rosen et al. \cite{Rosen2012}, found in Appendix \ref{appendix:RJMCMCsampling}. This scheme partitions the time series into locally stationary segments and models the log PSD $\log f(\mathbf{\nu})$  of the segments with a Gaussian process with covariance matrix $\Omega$. For computational savings, they employ an eigendecomposition $\Omega=QDQ^T$ and retain only the 10 largest eigenvectors in $D$. That is, they let $D_{10}$ be the truncated diagonal matrix and set $X=QD_{10}^{1/2}$ as their design matrix. Let $\beta$ be a vector of unknown regression coefficients with prior distribution $N(0,\tau^2 I)$ (where $\tau^2$ is a smoothing parameter). Then the successive estimate for the PSD is $ \log f(\mathbf{\nu})=X \mathbf{\beta}$, iterated within the RJMCMC.

Given that the eigendecomposition is so critical, the exact number of eigenvalues should be estimated using the data. Using a fixed number of 10 eigenvalues (as do Rosen et al., detailed above) yields a smoothing spline that may be too smooth depending on the process being estimated. If the true spectrum is highly smooth, our estimator will exhibit too much variance; if the spectrum is particularly complex, our estimator will exhibit excessive bias. Instead, we select an optimal number of eigenvalues $M$ by analysing posterior samples of the log spectrum for a range of eigenvectors, integrating over all possible values of regression coefficients, and maximising the marginal likelihood. We select our number of eigenvalues as follows:  
\begin{equation}
\label{eq:key defn eqn}
  \hat{M}= \argmax_{M \geq 1} \int p(\mathbf{y}|M, X_M, \beta) p(\mathbf{\beta} |M, X_M) d \mathbf{\beta},
\end{equation}
where $\mathbf{y}$ is the log periodogram data, $X_M$ is the design matrix under an eigendecomposition using $M$ eigenvalues, and $\mathbf{\beta}$ is a vector of coefficients determining the weight for respective eigenvectors. We refer to the corresponding eigendecomposition $\Omega=QD_{\hat{M}} Q^T$ as the \emph{optimal smoothing spline}.

We remark that the determination of this $\hat{M}$ requires post-processing after a sequence of independent MCMC runs, both in the stationary and nonstationary case, and in all implemented experiments throughout the paper. Specifically, $\hat{M}$ is not computed within any individual MCMC iteration, but by analyzing separate chains for each possible number of eigenvalues $M$.

We also remark that the value of $\hat{M}$ will generally be larger for more complex spectra, with a smaller number of eigenvectors being sufficient for relatively simple AR(1) and AR(2) processes. As for sample size, the relationship with $\hat{M}$ may be a little more complex. A larger length of the time series $n$ or $T$ may increase the complexity of the spectrum and hence $\hat{M}$, but not necessarily. For example, sampling a very large number of data points from a simple AR(1) process would not yield a higher optimal value of $\hat{M}$. In fact, a small sample size could lead to additional difficulty. If an autoregressive process is of insufficient length, it may be difficult to extract information about the underlying signal in the periodogram. This is particularly important for time series that would generate complex spectra such as the AR(4) processes outlined in this paper. A larger length of the time series will also lead to greater computational cost when computing the periodogram. This could be costly in a reversible jump procedure (especially when there are relatively few segments), as this computation must occur many times until the algorithm's termination.

In subsequent sections, we also explain the possibility of implementing a penalty on the number of eigenvalues. This would yield an alternative eigendecomposition we will refer to as the \emph{penalised optimal smoothing spline}.

In subsequent sections, we examine the performance of the optimal smoothing spline against both simulated and real data. First, we specialise to the stationary case, where we may validate this optimal smoothing spline in the case of autoregressive processes with known analytic spectra. In this setting, there is no need for the partition of the time series, so the RJMCMC scheme reduces to a Metropolis-Hastings scheme described in Appendix \ref{appendix:MH}. In this implementation, we are also free to substitute alternative GP covariance functions $k(\mathbf{x},\mathbf{x'})$ (see Definition \ref{def:GP} and Appendix \ref{appendix:MH}). The optimal smoothing spline compares favourably not only to the 10-eigenvector decomposition but also to the other covariance functions and provides further justification of its utility.

Having established that the optimal smoothing spline performs the best of our covariance functions in the stationary case, we proceed to a simulated nonstationary time series, once again using the RJMCMC. By examining three adjoined autoregressive processes with varying spectral complexity, we demonstrate the improvement in our method over the 10-eigenvector decomposition by validating each segment against its known analytic spectra. Finally, we proceed to test our optimal smoothing spline on real data, which is much less smooth than the autoregressive processes studied until this point. We show that the 10-eigenvector decomposition drastically underfits the many peaks of more complex real-world data. Whenever we use the reversible jump procedure to partition a nonstationary time series, we present our results conditional on the modal number of change points as a default.

\subsection{Penalised likelihood}
\label{sec:penalisation}

In this section, we discuss the possibility and potential advantages of imposing a penalty term in the selection of best number of eigenvalues $M$. Imposing a penalty term would inevitably select a smaller number of eigenvalues, and carries several advantages: 

\begin{enumerate}
    \item If desired, it would keep the estimate of the PSD smoother.
    \item This could reduce aliasing arising due to an insufficient number of data points.
    \item When we are uncertain of ideal model complexity, we may prefer simpler models.
    \item Different users may impose different costs of error. A user more sensitive to bias may favour the optimal, while a user more sensitive to variance may use a smaller number of eigenvalues.
\end{enumerate}

Imposing penalty terms is an intricate and complex question depending on the exact problem at hand. In this paper, we take an approach informed by observations of the behaviour of the marginal log likelihood against the number of eigenvalues. Observe subsequent plots of the log likelihood, both in synthetic experiments (Figures \ref{fig:Loglikelihood} and \ref{fig:NonStationarySpectrum}) and real data (Figure \ref{fig:SunspotsData}). Initially, an increasing number of eigenvectors provides a substantial (approximately) linear increase in the marginal log likelihood, followed by a clear ``elbow'' in the plots, where the rate of increase drastically reduces or levels off altogether. We wish to isolate this elbow using a penalised approach. Specifically, we want to impose a linear penalty term on the log marginal likelihood - this will help isolate the value of $M$ in which the growth of the log marginal likelihood changes from ``fast'' to ``slow.''

For this purpose, let $L(M)$ denote the log marginal likelihood
\begin{align}
\label{eq:loglikelihood}
    \mathcal{L}(M) = \log \int p(\mathbf{y}|M, X_M, \beta) p(\mathbf{\beta} |M, X_M) d \mathbf{\beta}.
\end{align}
Consider an interval of consideration, $[M_\text{min}, M_\text{max}].$ In our implementation, $M_\text{min}=10$, while $M_\text{max}$ is usually 70. Let 
\begin{align}
    \alpha = \frac{\mathcal{L}(M_\text{max}) - \mathcal{L}(M_\text{min})}{M_\text{max} - M_\text{min}}
\end{align}
be the ``overall gradient'' of the log likelihood curve. We impose the following penalty to determine the optimal penalised $\hat{M}_p$:
\begin{align}
\label{eq:penalisedM}
\hat{M}_p = \argmax_{M_\text{min} \leq M \leq M_\text{max} } \mathcal{L}(M) - \alpha M \\
= \argmax_{M_\text{min} \leq M \leq M_\text{max} }\log \left(\int p(\mathbf{y}|M, X_M, \beta) p(\mathbf{\beta} |M, X_M) d \mathbf{\beta}\right) \\  - \alpha M.
\label{eq:penalisedlikelihood}
\end{align}
Essentially, this penalised optimal $\hat{M}_p$ aims to select the value of $M$ at the elbow of the log marginal likelihood curve at ``the point'' (a heuristic) where (at least in the examples we observe) quick linear growth changes to much slower linear growth.

\section{Validation metrics}
\label{sec:validationmetrics}

In this section, we describe in detail both existing and new validation metrics - these will be used in the subsequent simulation study (Section \ref{simulation}) to compare our spectral density estimates ${\hat{\mathbf{g}}}$ of autoregressive processes with the known analytic log PSD ${\mathbf{g}}$. First, we outline the existing metrics we use, and then we describe in detail a new framework for measuring distances between spectral peaks, including the possibility of multiple non-trivial peaks in more complex spectra.

\subsection{Existing metrics}
\label{sec:existingmetrics}
The three existing metrics we use are root mean squared error (RMSE), defined as $\sqrt{\frac{1}{m} \sum_{j=0}^{m-1} (\hat{g}(\nu_j)-g((\nu_j))^2}$, mean absolute error (MAE), defined as $\frac{1}{m}\sum^{m-1}_{j=0}|\hat{g}(\nu_j)-g(\nu_j)|$, and the Wasserstein distance between the finite sets $\{g(\nu_j)\}$ and $\{\hat{g}(\nu_j)\}$. Specifically, this converts a finite set to a uniform probability measure over its elements and computes the Wasserstein distance between these measures. 

In further detail, let $S$ be a finite subset of $\mathbb{R}$. To $S$ we can associate a uniform measure defined as a weighted sum of Dirac delta measures
\begin{align}
\label{eq:Wasserstein delta}
    \mu_S=\frac{1}{|S|}\sum_{s \in S}\delta_s.
\end{align}
Integrating $\mu_S$ yields a cumulative density function $F$ and its associated quantile function $F^{-1}$. Concretely, if $S=\{s_1,...,s_n\} \subset \mathbb{R}$, then
\begin{align}
    F=\sum_{j=1}^{n-1} \frac{j}{n} \mathbbm{1}_{[s_j,s_{j+1})} + \mathbbm{1}_{[s_n,\infty)}, \\
    F^{-1}=\sum_{j=1}^{n} s_j \mathbbm{1}_{(\frac{j-1}{n},  \frac{j}{n})}.
    \label{eq:quantile}
\end{align}
Given two finite sets $S,T$, their $L^1$-Wasserstein distance is computed as $d_W(S,T)=W_1(\mu_S,\mu_T)$ and be computed as
\begin{align}
\label{eq:nextthing}
   d_W(S,T)=W_1(\mu_S,\mu_T)=\int_{0}^1 |F^{-1} - G^{-1}| dx,
\end{align}
where $F^{-1}$ and $G^{-1}$ are defined as in (\ref{eq:quantile}) from $S$ and $T$ respectively. For more detail, see \cite{james2021_crypto2}, Appendix C.

\subsection{Sets of spectral peaks and the proximity matching criterion}
\label{sec:multiplepeaks}

Our new validation metrics are inspired by applications in the physical sciences. There, a single dominant peak is often the greatest importance to observing scientists \cite{ToddChem}. Such a peak has two attributes, its amplitude, $\max \mathbf{g}$ and the frequency at which it occurs, $\argmax \mathbf{g}$. With this in mind, we introduce the simplest form of two validation metrics, $|\max \hat{\mathbf{g}}-\max \mathbf{g}|$ and $|\argmax \hat{\mathbf{g}}-\argmax \mathbf{g}|$, measuring how well the spectral density estimation determines the amplitude and frequency of the greatest peak.

Next, we consider the possibility of multiple peaks in an analytic or estimated power spectrum, as is the case for an AR(4) process. The remainder of this section is dedicated to carefully exploring this possibility, including how to measure distance between (non-trivial) sets of estimated and analytical spectral peaks.

First, we consider the simplest possible case, assuming the following \textit{a priori}:
\begin{enumerate}
    \item the analytic and estimated power spectra have the same number of peaks, $\rho_1,...,\rho_r$ for $\mathbf{g}$ and $\hat{\rho}_1,...,\hat{\rho}_r$ for $\hat{\mathbf{g}}$;
    \item there is a bijection between the two sets of peaks, for which each $\rho_i$ is the closest peak of $\mathbf{g}$ to $\hat{\rho}_i$, and vice versa.
\end{enumerate}
Informally, this is the ideal situation, in which the peaks of $\mathbf{g}$ can be matched one-to-one to their closest counterpart among the peaks of $\hat{\mathbf{g}}$, and vice versa. In this ideal case, we can define a validation metric between peaks' collective amplitudes and frequencies, respectively, for any $p\geq 1$ as follows:
\begin{align}
\label{eq:amplitudedist}
\left(\frac{|g(\rho_1) - \hat{g}(\hat{\rho}_1)|^p+...+|g(\rho_r) - \hat{g}(\hat{\rho}_r)|^p}{r}\right)^{\frac{1}{p}}, \\
\label{eq:reducedform} \left(\frac{|\rho_1 - \hat{\rho}_1|^p+...+|\rho_r - \hat{\rho}_r|^p}{r}\right)^{\frac{1}{p}}.
\end{align}
If $p=2$, we return a (normalised) Euclidean distance between tuples $(\rho_1,...,\rho_r)$ and $(\hat{\rho_1},...,\hat{\rho_r}) \in \mathbb{R}^r$. If $p=1$, we return an $L^1$ distance between the tuples. In subsequent paragraphs, we will see why the normalisation is appropriate for a more general framework.

More generally, we wish to outline a framework that specifically tests for whether this bijection exists, and define a validation measure that is well-defined regardless of whether the bijection exists. For this purpose, let $S$ be the finite set of the (non-trivial) peaks of an analytic log power spectrum $\mathbf{g}$, and $\hat{S}$ be the set of (non-trivial) peaks of an estimate $\hat{\mathbf{g}}$. The most naive definition of a peak as a local maximum of $\mathbf{g}$ or $\hat{\mathbf{g}}$, respectively, may be unsuitable for this purpose - due to the wiggly nature of spectral density estimates, there may exist local maxima of $\hat{g}$ that are insubstantial and should be excluded. An algorithmic framework for the refinement of peaks in the power spectra must be employed. A flexible algorithmic framework that does this for local maxima and minima is detailed in Appendix \ref{appendix:TPA}. For example, local maxima that are significantly less than the global maximum of the log PSD may be excluded as insubstantial under a variety of definitions and parameters. In our experiments with the AR(4) process, only a simple approach is needed: set $\delta=2$; any local maximum $\hat{\rho}$ of $\hat{\mathbf{g}}$ with $\hat{g}(\hat{\rho})<\max \hat{\mathbf{g}} - \delta$ is excluded. The same results are produced for any $\delta \in [2,4]$, demonstrating the robustness of this refinement procedure. As this is applied to the log power spectrum, such an inequality is preserved when scaling the initial time series $X_t$ by an affine transformation $X_t' = aX_t + b$. As an additional remark, we note that the precise number $r$ of (non-trivial) peaks is inevitably determined by the precise refinement algorithm we use. We discuss further details in Appendix \ref{appendix:TPA}.

Having identified such sets $S$ and $\hat{S}$, either through a choice of parameters within such a framework or by manual inspection, we define the \emph{proximity matching criterion} as follows: say $S$ and $\hat{S}$ satisfy this criterion if:
\begin{enumerate}
    \item there is a function $f:S \to \hat{S}$, where for each $s \in S$, $f(s)$ is the unique closest element of $\hat{S}$ to $s$; 
    \item there is a function $\hat{f}: \hat{S} \to S$, where conversely for each $\hat{s} \in \hat{S}$, $\hat{f}(\hat{s})$ is the unique closest element of $S$ to $\hat{s}$;
    \item The functions $f$ and $\hat{f}$ are mutually inverse.
\end{enumerate}
For example, suppose an analytic log PSD $\mathbf{g}$ has a set of non-trivial peaks $S=\{0.1, 0.2\}$. If a certain estimate $\hat{\mathbf{g}}_1$ has a set of non-trivial peaks $\hat{S}_1 = \{0.11, 0.19\}$ then there exists a bijection between $S$ and $\hat{S}_1$ that associates to each element of $S$ the closest element of $\hat{S}_1$, and vice versa. The bijection can be written down explicitly as $f:S \to \hat{S}_1, f(0.1)=0.11, f(0.2)=0.19$ and $\hat{f}: \hat{S}_1 \to S, \hat{f}(0.11)=0.1, \hat{f}(0.19)=0.2.$ Then, sets $S$ and $\hat{S}_1$ satisfy the proximity matching criterion.

For a second example, suppose an alternative estimate $\hat{\mathbf{g}}_2$ has a set of non-trivial peaks $\hat{S}_2=\{0.11, 0.12\}$. Then, there is no bijection that matches each element of $\hat{S}_2$ to the closest element on $S$, as both 0.11 and 0.12 are closest to the same element $0.1 \in S$. Thus no such proximity-matching bijection exists. For another example, if a third estimate  $\hat{\mathbf{g}}_3$ has a set of non-trivial peaks $\hat{S}_3=\{0.11\}$, then no bijection exists with $S$, so the criterion is not satisfied.

This criterion formalises the ideal case described above, where the (non-trivial) peaks of $\mathbf{g}$ and $\hat{\mathbf{g}}$ can be matched one-to-one and hence labelled $\rho_1,...,\rho_r$ and $\hat{\rho_1},...,\hat{\rho_r}$ via the mutually inverse bijective functions $f$ and $\hat{f}$. Simply put, the criterion formalises the notion that one can naturally pair up the non-trivial peaks of $\mathbf{g}$ and $\hat{\mathbf{g}}$ to facilitate a convenient comparison.

\subsection{Measures between sets of non-trivial peaks}
\label{sec:MJdefinition}

So far, we have defined a criterion that specifies the most convenient case, where one can conveniently measure distances between matched sets of peaks $S$ and $\hat{S}$ using (\ref{eq:amplitudedist}) and (\ref{eq:reducedform}). Finally, we define a measure between sets $S$ and $\hat{S}$ that makes sense in full generality, even when the proximity matching criterion fails. Again let $p \geq 1$; we adopt the \emph{semi-metrics} defined in \cite{James2019,James2021_geodesicWasserstein} and let
\begin{align}
\label{eq:MJ}
    d^p({S},\hat{S}) = \Bigg(\frac{\sum_{\hat{s}\in \hat{S}} d(\hat{s},S)^p}{2|\hat{S}|} + \frac{\sum_{{s} \in {S}} d(s,\hat{S})^p}{2|S|} \Bigg)^{\frac{1}{p}},
\end{align}
where $d(\hat{s},S)$ is the minimal distance from $\hat{s}$ to $S$, and vice versa. This validation semi-metric  can be applied to measure the collective distance between the sets of peaks of an analytic and estimated spectra under any circumstance, conditional on an appropriate selection of non-trivial peaks. (Such selection may be performed algorithmically or by inspection). Moreover, in the case where the proximity matching criterion is satisfied and the sets of peaks can be matched up and labelled as $\rho_1,...,\rho_r$ and $\hat{\rho_1},...,\hat{\rho_r}$, it simply reduces to to the metric (\ref{eq:reducedform}), together with its normalising factor, between tuples $(\rho_1,...,\rho_r)$ and $(\hat{\rho_1},...,\hat{\rho_r}) \in \mathbb{R}^r$.

To summarise, we have defined a general validation (semi)-metric between the collective sets of peaks of an analytic and estimated power spectra $\mathbf{g}$ and $\hat{\mathbf{g}}$, respectively. For full generality and precision, the sets of peaks must be selected from the data with a flexible algorithmic framework suitable for the application at hand, and not merely chosen by observation from the shape of the analytic spectrum. In the most suitable case, this semi-metric will reduce to what is still a new validation metric defined in Eq. (\ref{eq:reducedform}).

\section{Simulation study}
\label{simulation}
In this section, we study the properties of our estimators with a detailed simulation study. We simulate three stationary autoregressive processes, AR(1), AR(2) and AR(4), and validate our spectral density estimates against their known analytic power spectra. For each experiment, we run 50 simulations with our MCMC scheme, either RJMCMC or Metropolis-Hastings, as described in Section \ref{sec:model and priors}.  Each sampling scheme consists of 10 000 iterations, with 5000 used for burn-in, and each process contains $n = 500$ data points. Following the notation of Section \ref{sec:model and priors}, we set $m=251$ and consider Fourier frequencies $\nu_j=\frac{j}{n}, j=0,1,...,m-1$, the appropriate range of the periodogram and spectrum. As such, the range of the frequency axis in all plots is $0\leq \nu \leq \frac{1}{2}.$ Whenever we report results from our validation metrics (described in Section \ref{sec:validationmetrics}), we always average results across the 50 performed simulations. Henceforth, $\epsilon_t$ denotes a white noise random variable with distribution $N(0,1)$.

\begin{figure*}
    \centering
    \begin{subfigure}[b]{0.3\textwidth}
        \includegraphics[width=\textwidth]{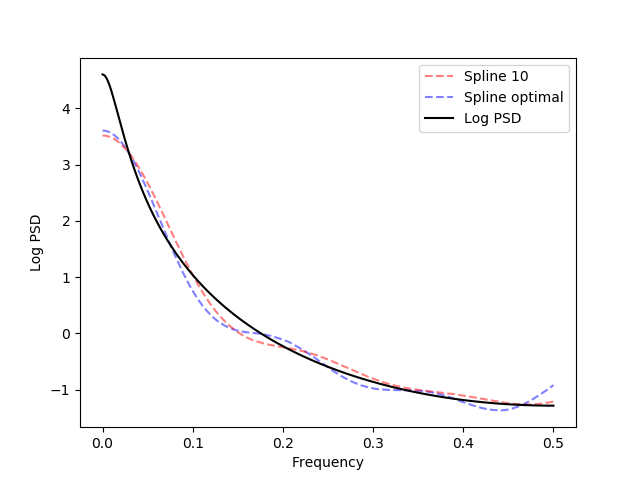}
        \caption{}
        \label{fig:Spline_AR1_likelihood}
    \end{subfigure}
    \begin{subfigure}[b]{0.3\textwidth}
        \includegraphics[width=\textwidth]{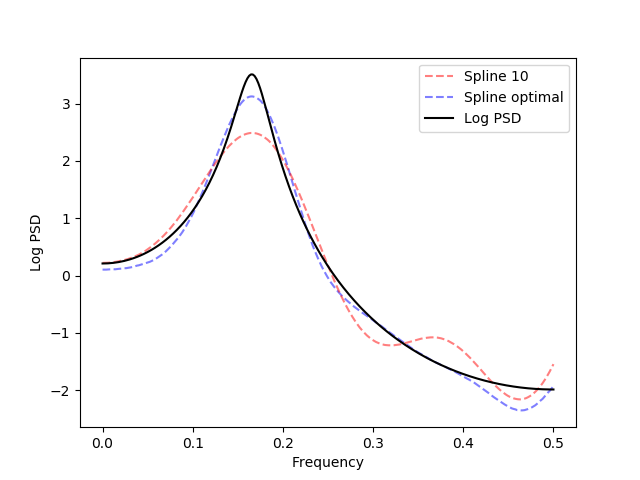}
        \caption{}
        \label{fig:Spline_AR2_likelihood}
    \end{subfigure}
    \begin{subfigure}[b]{0.3\textwidth}
        \includegraphics[width=\textwidth]{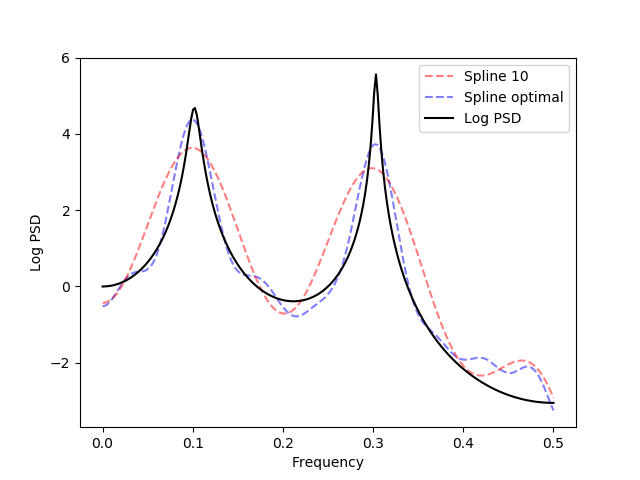}
        \caption{}
        \label{fig:Spline_AR4_likelihood}
    \end{subfigure}
     \begin{subfigure}[b]{0.3\textwidth}
        \includegraphics[width=\textwidth]{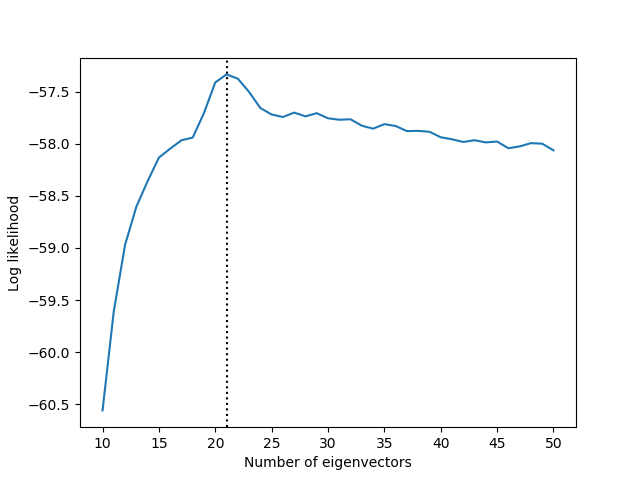}
        \caption{}
        \label{fig:Likelihood_Spline_AR1_likelihood}
    \end{subfigure}
    \begin{subfigure}[b]{0.3\textwidth}
        \includegraphics[width=\textwidth]{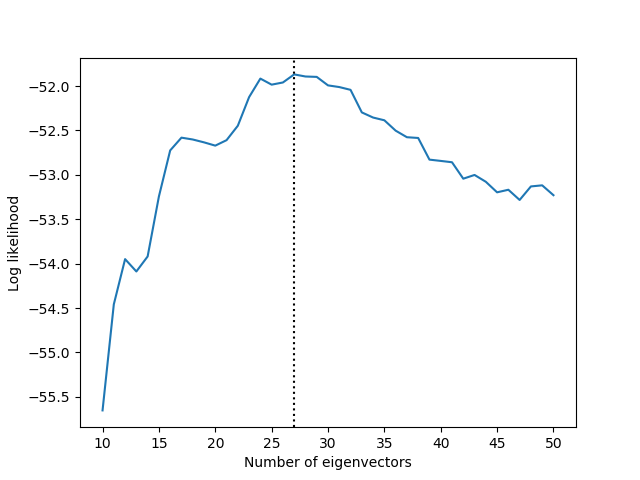}
        \caption{}
        \label{fig:Likelihood_Spline_AR2_likelihood}
    \end{subfigure}
    \begin{subfigure}[b]{0.3\textwidth}
        \includegraphics[width=\textwidth]{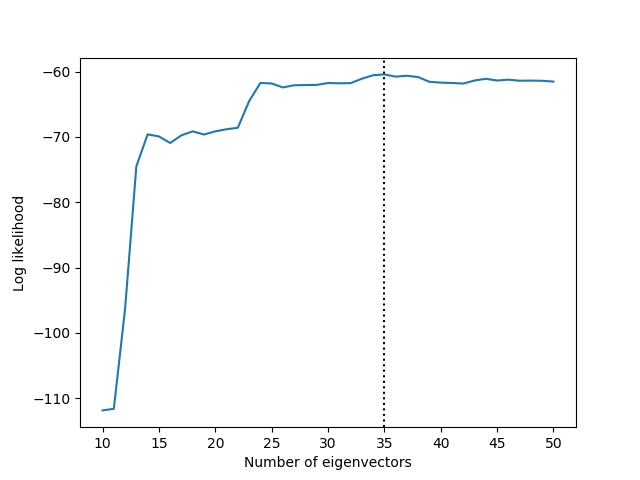}
        \caption{}
        \label{fig:Likelihood_Spline_AR4_likelihood}
    \end{subfigure}
    \caption{Analytic and estimated log PSD for 10-eigenvector and optimal decomposition of the smoothing spline models for (a) AR(1) (b) AR(2) (c) AR(4). Log likelihood vs number of eigenvectors for (d) AR(1) (e) AR(2) and (f) AR(4) process. The peak in the log likelihood determines the number of eigenvectors in the optimal model.}
    \label{fig:Loglikelihood}
\end{figure*}

\begin{table*}
\centering
\begin{tabular}{ |p{3.25cm}||p{1cm}|p{.8cm}|p{2cm}|p{3.5cm}|p{3cm}|  }
 \hline
 \multicolumn{6}{|c|}{AR(1) model} \\
 \hline
 Covariance function & RMSE & MAE & Wasserstein & $|\argmax g - \argmax \hat{g}|$ & $|\max g - \max \hat{g} |$ \\
 \hline
 Spline$_{10}$ & 0.26    & 0.20 & 0.18 & 0 & 0.90  \\
 Spline$_\text{optimal}$ & 0.25   & 0.19 & 0.16 & 0 & 0.93 \\
 Squared exponential  & 0.16  & 0.12 & 0.12 & 0 & 0.49 \\
 Matern$_{3/2}$ & 0.19 & 0.15 & 0.13  & 0 & 0.51 \\
 Matern$_{5/2}$ & 0.19 & 0.14 & 0.12 & 0 & 0.49\\
 Sigmoid & 0.15 & 0.11 & 0.11  & 0 & 0.34\\
 SQE+Sigmoid & 0.18 & 0.14 & 0.13 & 0 & 0.40 \\
 \hline	
 \multicolumn{6}{|c|}{AR(2) model} \\
 \hline
 Covariance function & RMSE & MAE & Wasserstein & $|\argmax g - \argmax \hat{g}|$ & $|\max g - \max \hat{g} |$ \\
 \hline
 Spline$_{10}$ & 0.33 & 0.26 & 0.24  & 0.0020 & 0.73 \\
 Spline$_\text{optimal}$ & 0.21   & 0.17 & 0.14 & 0.0020 & 0.36 \\ 
 Squared exponential & 0.22 & 0.17 & 0.15 & 0.0020 & 0.56 \\
 Matern$_{3/2}$ & 0.21 & 0.17 & 0.14 & 0.0020 & 0.24 \\
 Matern$_{5/2}$ & 0.19 & 0.15 & 0.13 & 0.0020 & 0.37 \\
 Sigmoid & 0.32 & 0.24 & 0.22 & 0.0040  & 0.85 \\
 SQE+Sigmoid & 0.21 & 0.17 & 0.15  & 0.0020 & 0.53 \\
 \hline	
  \multicolumn{6}{|c|}{AR(4) model} \\
 \hline
 Covariance function & RMSE & MAE & Wasserstein & $|\argmax g - \argmax \hat{g}|$ & $|\max g - \max \hat{g} |$ \\
 \hline
 Spline$_{10}$ & 0.82 & 0.63 & 0.54 & 0.20 & 2.23 \\
 Spline$_\text{optimal}$ & 0.43   & 0.33 & 0.26 & 0.20 & 1.57 \\
 Squared exponential & 0.45  & 0.31 & 0.27 & 0.20 & 1.98\\
 Matern$_{3/2}$ & 0.37 & 0.27 & 0.21  & 0.20 & 1.75 \\
 Matern$_{3/2}$ & 0.40 & 0.29 & 0.24 & 0.20 & 1.90 \\
 Sigmoid & 0.97 & 0.75 & 0.41 & 0.21 & 2.12 \\ 
 SQE+Sigmoid & 0.45 & 0.32 & 0.26 & 0.20 & 2.06 \\ 
 \hline	
\end{tabular}
\caption{Results for synthetic data experiments. Mean error over 50 simulations is recorded for various covariance functions and validation metrics. Each MCMC scheme consists of 10,000 iterations where 5,000 are used for burn-in. For each AR process, $n=1000$.}
\label{tab:result_table}
\end{table*}

First, we generate data from an AR(1) process $x_{t} = 0.9 x_{t-1} + \epsilon_{t}$. We compute the log periodogram from the observed data and estimate the log PSD and GP hyperparameters with our MCMC scheme. Table \ref{tab:result_table} indicates that the best performing covariance functions are the squared exponential and Sigmoid.  The log PSD has high power at low frequency and gradually declines in power when moving toward higher frequency components. The spectrum does not exhibit any sharp peaks and it is unsurprising that the highly smooth squared exponential performs well in estimating the log PSD. The worst performing covariance function is the 10-eigenvector decomposition of the smoothing spline of Rosen et al \cite{Rosen2012}. There is little improvement with the optimal smoothing spline model in this case, as seen in Figure \ref{fig:Spline_AR1_likelihood}. This experiment suggests that for a simple spectrum, there is limited improvement in optimising the number of basis functions in the smoothing spline and alternative GP covariance functions provide superior performance. 

Next, we generate data from an AR(2) process $x_{t} = 0.9 x_{t-1} - 0.8 x_{t-2} + \epsilon_{t}$, compute the log periodogram and estimate the log PSD and GP hyperparameters with the MCMC schemes. Table \ref{tab:result_table} indicates that several covariance functions perform similarly well: SQE+Sigmoid, squared exponential, both Matern functions, and the optimal spline. The Sigmoid kernel and 10-eigenvector decomposition of the spline perform worse. The log PSD has a relatively flat gradient and a mild peak at $\nu \sim 0.15$. Once again, smoother covariance functions tend to perform better, while the Sigmoid covariance function, popularised for its ability to model abrupt changes in data modelling problems \cite{Rasmussen2006}, is a poor performer. Figure \ref{fig:Spline_AR2_likelihood} demonstrates a substantial improvement in estimating the log PSD when an optimal number, larger than 10, of basis functions is used. In particular, the estimator does a better job of detecting the maximum amplitude of the power spectrum. 

Third, we follow \cite{Edwards2018}, generate data from an AR(4) process $x_{t} = 0.9 x_{t-1} - 0.9 x_{t-2} + 0.9 x_{t-3} - 0.9 x_{t-4} + \epsilon_{t}$, and again estimate the log PSD and GP hyperparameters with the MCMC schemes. The log PSD of the AR(4) process is more complex than the prior two experiments. There are two peaks of different amplitudes in the spectrum, and the spectrum changes more abruptly than the AR(1) and AR(2) spectra. Table \ref{tab:result_table} indicates that the best performing covariance functions are the optimal smoothing spline and the two Matern functions. The worst performing covariance functions are the 10-eigenvector smoothing spline and the Sigmoid kernel. Once again we see a substantial improvement in estimation when an optimal number of basis functions is used, displayed in Figure \ref{fig:Spline_AR4_likelihood}.

As the AR(4) process has two peaks, we may also test the proximity matching criterion of Section \ref{sec:multiplepeaks} and measure the distance between the collective sets of peaks of the estimated and analytic power spectra. In Table \ref{tab:MJ1_result_table}, we note that in this relatively simple example with just two peaks, all covariance functions satisfy this criterion. As such, the more general semi-metric (\ref{eq:MJ}) reduces to the simpler metric between frequencies (\ref{eq:reducedform}). We also include the collective distance between amplitudes in the table, as defined in (\ref{eq:amplitudedist}). These distances show that the optimal smoothing spline does the best job at estimating both peaks at the same time.

Figures \ref{fig:Likelihood_Spline_AR1_likelihood}, \ref{fig:Likelihood_Spline_AR2_likelihood} and \ref{fig:Likelihood_Spline_AR4_likelihood} plot the Whittle likelihood against the number of eigenvectors for the AR(1), AR(2) and AR(4) processes, respectively - this selects the optimal number of basis functions for each. We see that consistently more than 10 basis functions are required to optimally model the data; in fact, the choice of 10 basis functions in the eigendecomposition is only really appropriate for the AR(1) and AR(2) spectrum. Even the synthetic AR(4) spectrum, but more so real-world processes, benefit from an optimised decomposition that can better suit their complexity and lack of smoothness. Indeed, real-world time series may have multiple periodic components represented in complex power spectra with multiple peaks, of varying amplitudes. The fixed 10-eigenvector decomposition is excessively smooth for these more complex spectra. In Appendix \ref{appendix:PW}, we examine an alternative AR(4) process, known in the literature for its difficulties.

\begin{table*}
\centering
\begin{tabular}{|p{3.25cm}|p{3cm}|}
 \hline
 \multicolumn{2}{|c|}{AR(4) process: distance between sets of peaks} \\
 \hline
 Covariance function  & Amplitude distance \\
 \hline
 Spline$_{10}$  & 1.64  \\
 Spline$_\text{optimal}$  & 0.99 \\
 Squared exponential   & 1.24\\
 Matern$_{3/2}$  & 1.18 \\
 Matern$_{5/2}$  & 1.31\\
 Sigmoid kernel  & 1.57 \\
 SQE+Sigmoid  & 1.34 \\
 \hline	
\end{tabular}
\caption{Amplitude distance between collective sets of peaks of analytic and estimated AR(4) power spectra $g$ and $\hat{g}$, respectively. The proximity matching criterion is satisfied in every instance.}
\label{tab:MJ1_result_table}
\end{table*}

\begin{table*}
\centering
\begin{tabular}{ |p{3.25cm}||p{1cm}|p{.8cm}|p{2cm}|p{3.5cm}|p{3cm}|  }
 \hline
 \multicolumn{6}{|c|}{Piecewise autoregressive model} \\
 \hline
Method & RMSE & MAE & Wasserstein & $|\argmax g - \argmax \hat{g}|$ & $|\max g - \max \hat{g} |$ \\
 \hline
 Spline$_{10}$ & 2.81    & 2.04 & 1.65 & 0.21 & 6.81  \\
 Spline$_\text{optimal}$ & 1.32   & 1.01 & 0.58 & 0.21 & 5.21 \\
  Spline$_\text{penalised optimal}$ & 1.33   & 0.96 & 0.70 & 0.20 & 6.22 \\
 \hline	
\end{tabular}
\caption{Results for experiment on synthetic nonstationary time series described in Eq. (\ref{eq:simulationpiecewisestationary}). Each validation metric is obtained by averaging the validation metrics on each segment, comparing the estimated log power spectra to the analytic known spectra. The optimal number of eigenvalues is 61, the penalised optimal is 29.}
\label{tab:nonstationarymetrics}
\end{table*}

\begin{figure*}
    \centering
    \begin{subfigure}[b]{0.40\textwidth}
        \includegraphics[width=\textwidth]{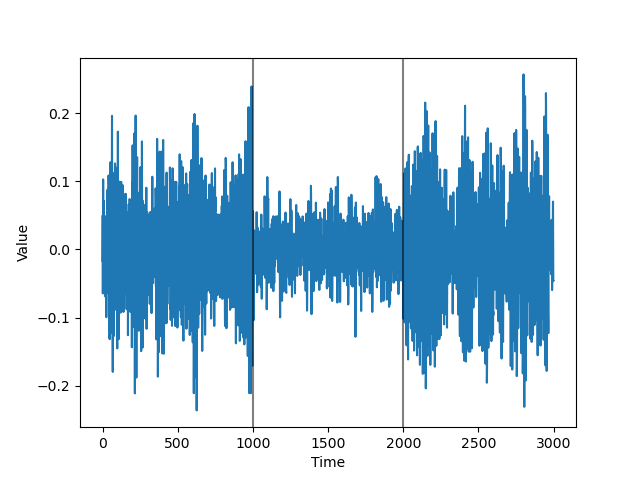}
        \caption{}
        \label{fig:NonstationaryRealisation}
    \end{subfigure}
    \begin{subfigure}[b]{0.40\textwidth}
        \includegraphics[width=\textwidth]{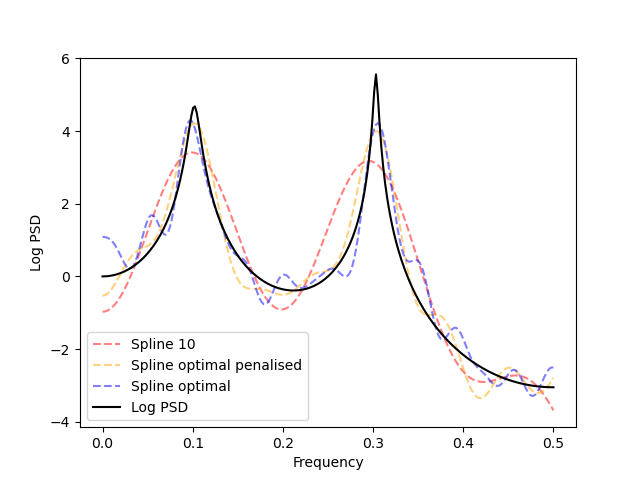}
        \caption{}
        \label{fig:Segment1}
    \end{subfigure}
    \begin{subfigure}[b]{0.40\textwidth}
        \includegraphics[width=\textwidth]{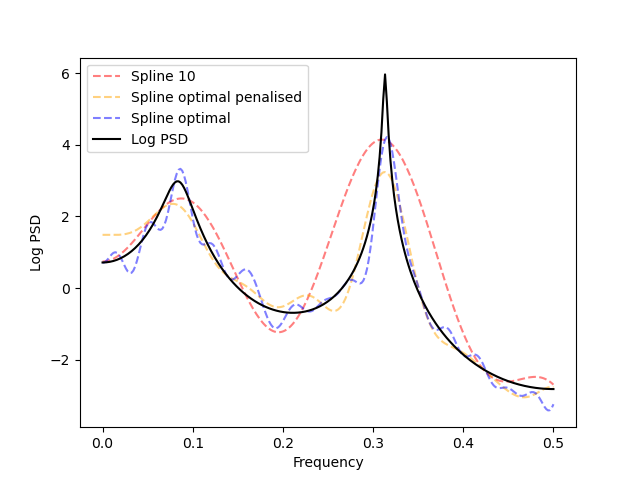}
        \caption{}
        \label{fig:Segment2}
    \end{subfigure}
    \begin{subfigure}[b]{0.40\textwidth}
        \includegraphics[width=\textwidth]{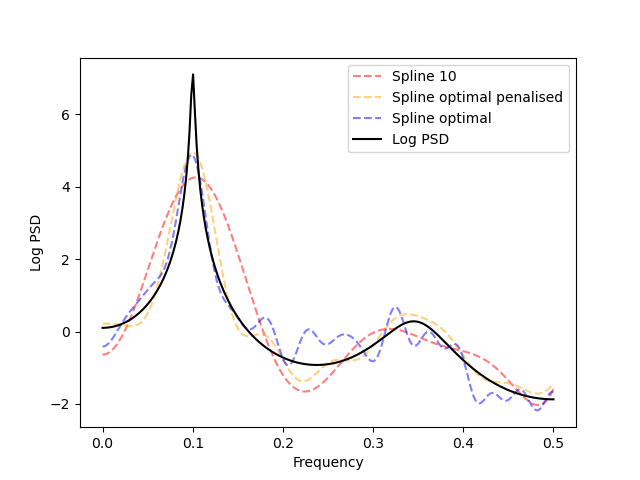}
        \caption{}
        \label{fig:Segment3}
    \end{subfigure}
    \begin{subfigure}[b]{0.40\textwidth}
        \includegraphics[width=\textwidth]{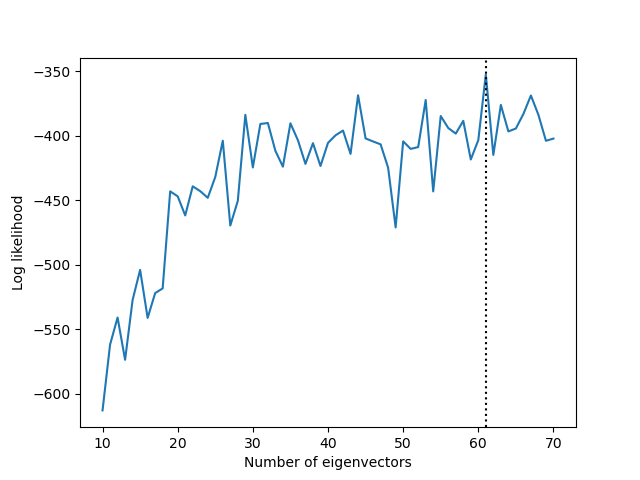}
        \caption{}
        \label{fig:NonstationaryLikelihood}
    \end{subfigure}
    \begin{subfigure}[b]{0.40\textwidth}
        \includegraphics[width=\textwidth]{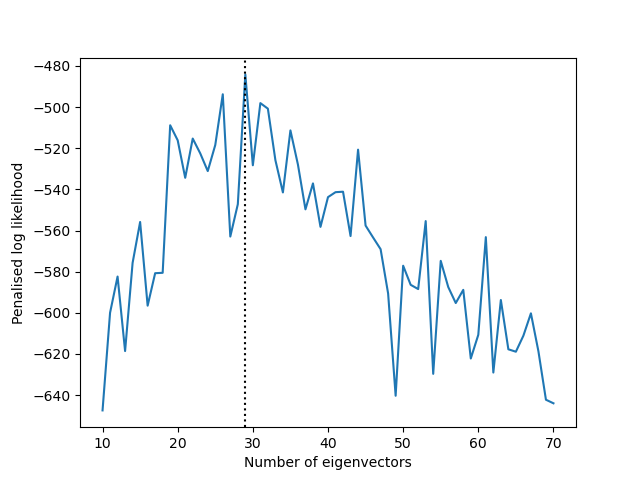}
        \caption{}
        \label{fig:NonstationaryLikelihoodpenalised}
    \end{subfigure}
      \begin{subfigure}[b]{0.40\textwidth}
        \includegraphics[width=\textwidth]{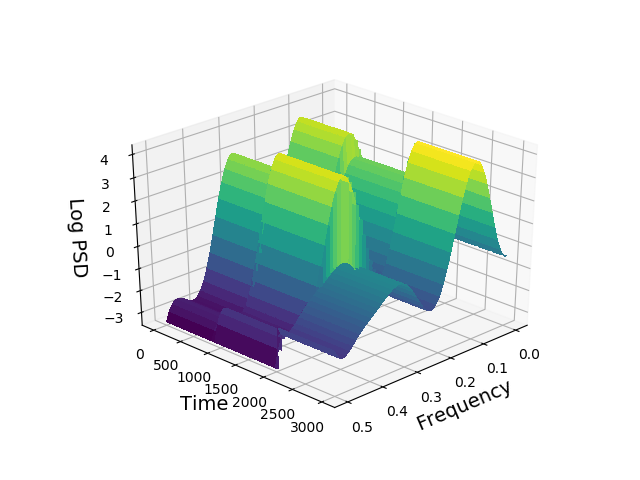}
        \caption{}
        \label{fig:TVS_10}
    \end{subfigure}
    \begin{subfigure}[b]{0.40\textwidth}
        \includegraphics[width=\textwidth]{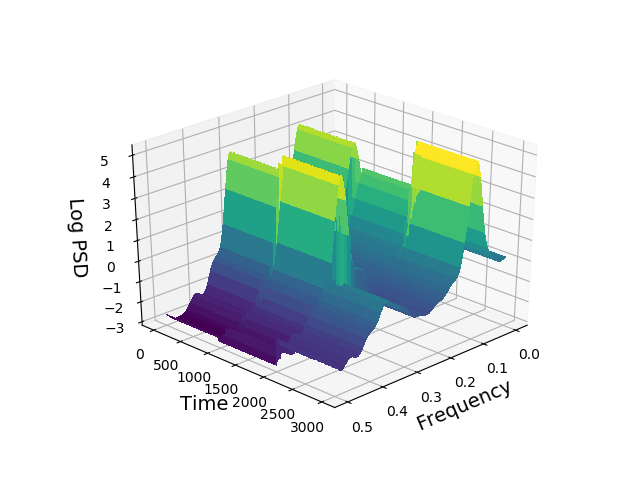}
        \caption{}
        \label{fig:TVS_optimal}
    \end{subfigure}
    \caption{Analytic and estimated log PSD for 10-eigenvector and optimal decomposition of smoothing spline models for nonstationary time series. (a) Realisation of the time series. Spectral estimates for (b) Segment 1 (c) Segment 2 (d) Segment 3, each using the optimal and penalised optimal spline. (e) Plot of log marginal likelihood vs number of eigenvectors, using Eq. (\ref{eq:loglikelihood}). The optimal number of eigenvectors is 61. (f) Plot of the penalised log likelihood function, using Eq. (\ref{eq:penalisedM}). The penalised optimal number of eigenvectors is 29.  Spectral estimates for  Spectral surface (time-varying PSD) for (g) 10-eigenvector and (h) optimal eigendecomposition.}
    \label{fig:NonStationarySpectrum}
\end{figure*}

Next, we include a simulation study to demonstrate the efficacy of our improved method at both segmenting a nonstationary time series and then estimating the power spectrum of each locally stationary segment. We generate a piecewise autoregressive time series by concatenating three AR processes of segment length $n=1000$:
\begin{align}
\label{eq:simulationpiecewisestationary}
x_{t}=\begin{cases}
			0.9 x_{t-1} - 0.9 x_{t-2} + 0.9 x_{t-3} - 0.9 x_{t-4} + \epsilon_{t}, & \\ \text{1 $\leq t \leq$ 1000};\\
            0.8 x_{t-1} - 0.6 x_{t-2} + 0.9 x_{t-3} - 0.8 x_{t-4} + \epsilon_{t}, & \\ \text{1001  $\leq t \leq$ 2000}; \\
            0.7 x_{t-1} - 0.15 x_{t-2} + 0.1 x_{t-3} - 0.6 x_{t-4} + \epsilon_{t}, & \\ \text{2001 $\leq t \leq$ 3000}.
		 \end{cases}
\end{align}
A realisation of this process is displayed in Figure \ref{fig:NonstationaryRealisation}. We remark that this is a more complex synthetic time series than that simulated by Rosen et al. (Figure 1, \cite{Rosen2012}) in which they concatenate AR(1) and AR(2) processes. Indeed, an important part of the improvement of our methodology is the fact that their method suffers from limitations when the underlying spectra are more complex than that of AR(1) or AR(2) processes, while our method has the flexibility to estimate the complex spectra of more involved processes and real-world data.

As before, we plot the Whittle likelihood against the number of eigenvectors in Figure \ref{fig:NonstationaryLikelihood}  - this selects the optimal number of basis functions for the optimal smoothing spline. We remark that the optimal number of basis functions is computed by optimising over the entire time-varying spectral surface. In this experiment, we also plot the penalised marginal likelihood function defined by (\ref{eq:penalisedM}) in Figure \ref{fig:NonstationaryLikelihoodpenalised}. These two figures clearly demonstrate the need for optimising the number of eigenvectors. Indeed, the 10-eigenvector decomposition provides the lowest likelihood of the data, while 61 eigenvectors are optimal and 29 are optimal in the penalised framework. Figures \ref{fig:Segment1}, \ref{fig:Segment2} and \ref{fig:Segment3} show that the optimal and penalised optimal smoothing spline models provide improved performance over each segment relative to the benchmark model of Rosen et al.

First, we report the posterior distributions of the number and locations of the change points. The three-segment model ($m=3$) is selected with probability 1. The first change point (corresponding to the synthetic break at $t=1000$) is estimated to have the following distribution: $t=1032$ with probability 0.39, $t=1035$ with probability 0.53, $t=1070$ with probability 0.08. The second change point is estimated to be $t=2002$ with probability 1.

In Figure \ref{fig:Segment1}, the optimal smoothing spline more effectively captures abrupt peaks in the PSD. We remark that the slight overfitting exhibited by the optimal spline would not lead to erroneous inference, as this overfitting only occurs at frequency components with limited power. In Figure \ref{fig:Segment2}, the optimal spline does a superior job at estimating both peaks in amplitude, while avoiding overfitting at frequency components that exhibit the most power. In Figure \ref{fig:Segment3}, we see two peaks in the PSD with significantly varying amplitude. There, the optimal model more appropriately determines the amplitude and frequency of the greatest peaks, just like our validation metrics in Table \ref{tab:result_table} recorded. 

Figures \ref{fig:TVS_10} and \ref{fig:TVS_optimal} display the two time-varying PSD estimated using the original 10-eigenvector decomposition and our optimal spline, respectively. Figure \ref{fig:TVS_optimal} clearly generates a more intricate surface than Figure \ref{fig:TVS_10}, which is inappropriately smooth. In particular, the scale of the $z$-axis demonstrates the optimised model's improvement when estimating appropriate peaks at respective frequencies in the PSD.

\begin{figure*}
    \centering
    \includegraphics{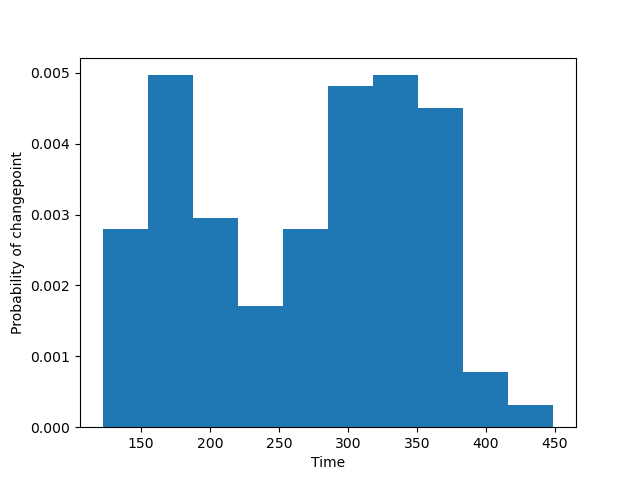}
    \caption{Estimate for the posterior distribution for a slowly varying autoregressive process defined in (\ref{eq:timevaryingAR}). The modal number of change points is 1, and conditional on this, the location of the change point is approximately uniform, subject to minimum bounds on each segment length.}
    \label{fig:slowlyvarying}
\end{figure*}

Finally, we include an alternative synthetic experiment, where an autoregressive process slowly varies over time, rather than exhibiting clear synthetic breaks. We consider a $T=500$ length time series examined by Rosen et al, defined by
\begin{align}
\label{eq:timevaryingAR}
    x_t= (-0.5 + \frac{t}{T})x_{t-1} + \epsilon_t
\end{align}
In this instance, the optimal and penalised optimal number of eigenvalues is $M=10$, as expected for a relatively simple (time-varying) AR(1) process. The modal number of segments is $m=2$. Conditional on two segments (that is, one change point), the estimated posterior distribution of this change point is displayed in Figure \ref{fig:slowlyvarying}. As expected for a process with no clear change point, the posterior distribution is approximately uniform, subject to the constraints of a minimal possible segment length.

\section{Real data}
\label{real data}
\begin{figure*}
    \centering
    \begin{subfigure}[b]{0.48\textwidth}
        \includegraphics[width=\textwidth]{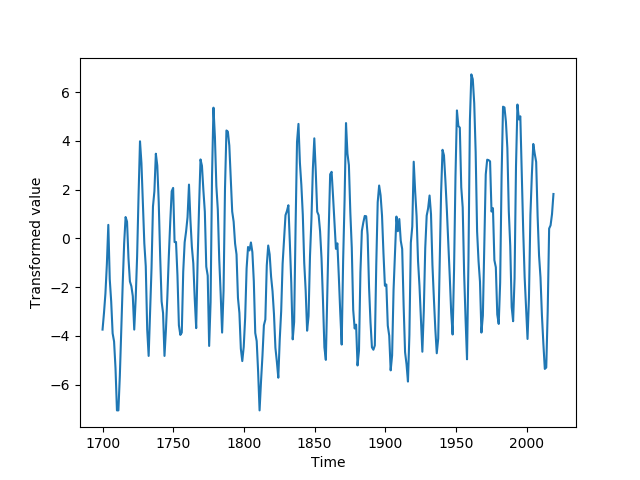}
        \caption{}
        \label{fig:Sunspot_transformed_time}
    \end{subfigure}
    \begin{subfigure}[b]{0.48\textwidth}
        \includegraphics[width=\textwidth]{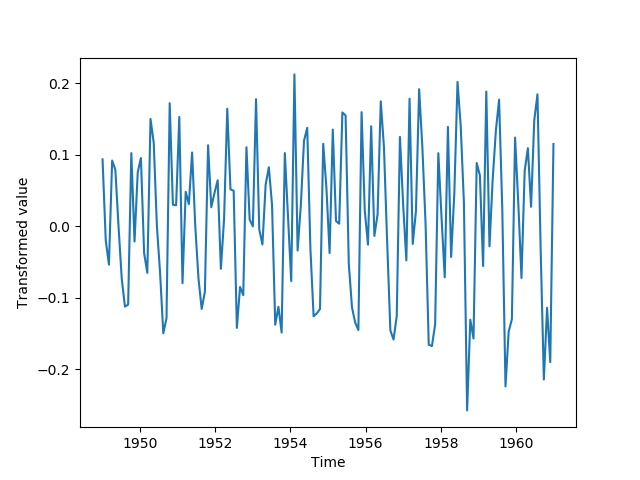}
        \caption{}
        \label{fig:Airline_transformed_time}
    \end{subfigure}
    \begin{subfigure}[b]{0.48\textwidth}
        \includegraphics[width=\textwidth]{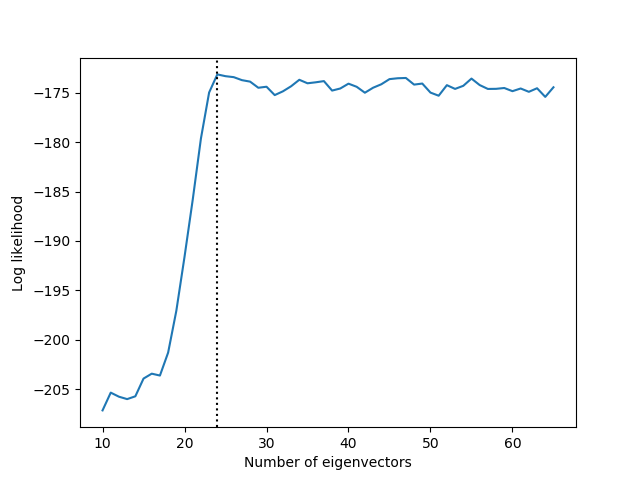}
        \caption{}
        \label{fig:EigenvectorsSunspot}
    \end{subfigure}
    \begin{subfigure}[b]{0.48\textwidth}
        \includegraphics[width=\textwidth]{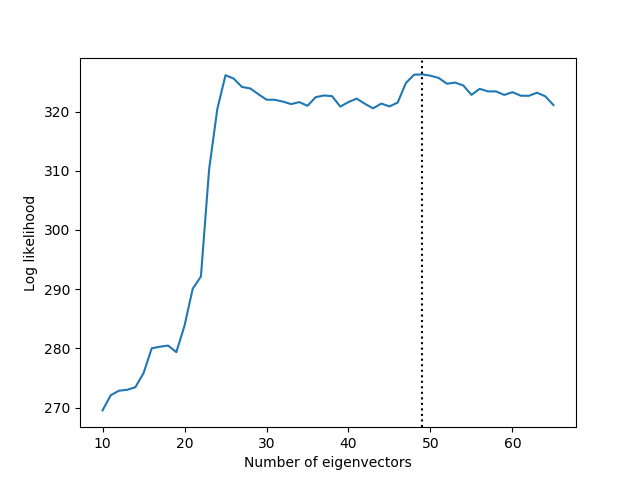}
        \caption{}
        \label{fig:EigenvectorsAirline}
    \end{subfigure}
    \begin{subfigure}[b]{0.48\textwidth}
        \includegraphics[width=\textwidth]{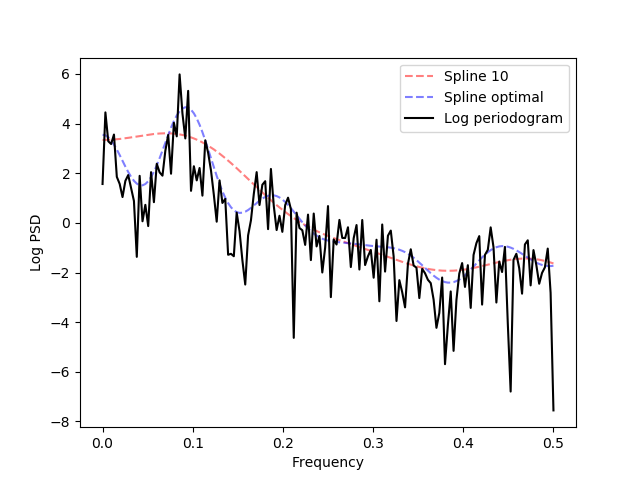}
        \caption{}
        \label{fig:SunspotsSpectrum}
    \end{subfigure}
    \begin{subfigure}[b]{0.48\textwidth}
        \includegraphics[width=\textwidth]{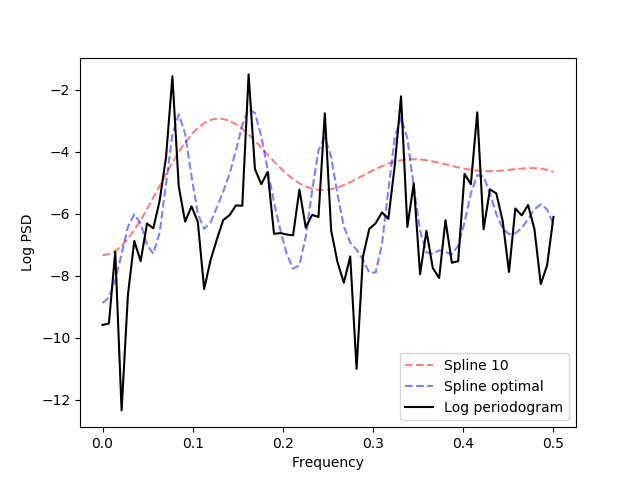}
        \caption{}
        \label{fig:AirlineSpectrum}
    \end{subfigure}
    \caption{Transformed time series for (a) sunspots (b) airline passenger data. Log likelihood vs number of eigenvectors for (c) sunspots (d) airline passenger data. Estimated log PSD for 10-eigenvector and optimal decomposition smoothing spline models for (e) sunspots (f) airline passenger data.}
    \label{fig:SunspotsData}
\end{figure*}

In this section, we analyse the spectra of two canonical datasets, sunspots \cite{Choudhuri2004} and airline passenger data \cite{Boxbook}. Both time series are well studied, with the former being particularly well studied in the spectral density estimation literature. We compare the performance of the optimal smoothing spline model and the 10-eigenvector smoothing spline. Unlike AR processes, real data do not possess analytic power spectra to validate estimates against. Both time series require transformations to ensure they are stationary. For the sunspots data, following Choudhuri \cite{Choudhuri2004}, we perform the following transformation:
$$ S_t = y^{\frac{1}{2}}_t - \mu(\{y^{\frac{1}{2}}_t:t=1,...,n \}). $$
That is, we take the square root of the observations and then mean-centre the resulting values. For the airline passenger data, we perform an original transformation based on first differences of fourth roots:
$$A_t = y_t^{\frac{1}{4}} - y_{t-1}^{\frac{1}{4}}.$$
 The two transformed time series are displayed in Figures \ref{fig:Sunspot_transformed_time} and \ref{fig:Airline_transformed_time}. As before, the number of basis functions is optimised to maximise the marginal Whittle likelihood. For the sunspots and airline passenger data respectively, the optimal number of basis functions is 24 and 49, seen in Figures \ref{fig:EigenvectorsSunspot} and \ref{fig:EigenvectorsAirline} respectively. The modal number of change points in both instances is 1, which is fitting as the transformed time series are approximately stationary. Thus, applying the stationary or nonstationary model from Section \ref{sec:model and priors} yields identical results. The priors we use are detailed in that section.

Figures \ref{fig:SunspotsSpectrum} and \ref{fig:AirlineSpectrum} demonstrate that our improved method models these complex processes substantially better than the existing 10-eigenvector model of Rosen et al. \cite{Rosen2012}, and that the latter does not have the complexity to model such processes. In Figure \ref{fig:SunspotsSpectrum}, the optimal smoothing spline captures the abruptness and amplitude of the one significant peak, as well as most of the movement of the log periodogram, while removing some of its characteristic noise. The 10-eigenvector spline fails to capture these features, with only slight recognition of that early peak. In Figure \ref{fig:AirlineSpectrum}, the optimal smoothing spline captures all five peaks of this spectrum and most of its complex undulating behaviour; the 10-eigenvector spline fails to capture the prominent peaks in the data or provide any meaningful or accurate inference of the spectrum.

\section{Conclusion}
\label{conclusion}

This paper has proposed and demonstrated the use of an improved algorithm for spectral density estimation of stationary and nonstationary time series. In addition to performing better quantitatively with respect to new and existing metrics, we have shown significantly better performance at appropriately matching the abruptness and complexity present in the PSD of real-world processes. The existing 10-eigenvector decomposition of Rosen et al. \cite{Rosen2012} appropriately models smooth processes like the AR(1) and AR(2), but substantially underfits the real data we have presented; our optimal smoothing spline has a greater capability to fit both smooth and abruptly changing spectra.

As further justification of the utility of the optimised eigendecomposition, we examine alternative covariance functions other than the smoothing spline and compare these against both the optimal and existing spline model in several settings. Our simulation study confirms that most well-known covariance functions perform acceptably well when analysing the power spectral density for stationary time series. Generally, smoother covariance functions such as the squared exponential and Matern perform better in the cases of smoother and simpler spectra, while models that combine stationary and nonstationary covariance functions such as the SQE+Sigmoid appear to generalise best across more complex spectral densities. The 10-eigenvector decomposition of the smoothing spline, as used within \cite{Rosen2012}, is the worst performing model. In particular, we demonstrate that eigenvector optimisation is what constitutes improvement in this case, not alternative covariance functions. Simply put, given that the eigendecomposition is so critical, it really should be learned from the data.

Our paper is not without limitations: first, while this paper improves upon the existing work of Rosen et al. \cite{Rosen2012}, there is no change to the underlying transdimensional sampling. Changing the transdimensional part of the model may lead to numerous further complications. If we were to make the algorithm locally adaptive in different segments of the time series, this could lead to an explosion in the number of model parameters and hence the computational cost of the procedure. In addition, allowing different numbers of eigenvalues in each segment could complicate the priors and likelihood computations that are involved in determining the segments themselves.

There are a variety of interesting directions for future research. First, future work could explore computationally feasible ways of creating a locally adaptive framework for PSD estimation in nonstationary time series, despite the challenges described above. That is, one could conceivably partition a locally stationary time series while at the same time optimising the number of basis functions used in the MCMC scheme within each segment, rather than over the entire time series period, as we have done. Future modelling frameworks could consider Bayesian and frequentist methods for identifying and learning the most appropriate covariance functions, or combinations thereof, to use within each segment of the time series. This could be performed within a Bayesian or non-Bayesian framework. We could also combine the eigendecomposition with the other covariance functions we have explored, including a learned choice of best covariance function through model averaging. Finally, our new validation metrics focused on collective sets of peaks, including the proximity matching criterion and algorithms for the appropriate refinement of peaks, could be employed in other contexts. These could be used together to measure a model's effectiveness at specifically identifying a set of multiple peaks arising from more complex data.

\begin{acknowledgements}
Many thanks to Lamiae Azizi, Sally Cripps and Alex Judge for helpful discussions.
\end{acknowledgements}

 \section*{Conflict of interest}
The authors declare that they have no conflict of interest.

\appendix

\section{Percival-Walden AR process}
\label{appendix:PW}

\begin{figure*}
    \centering
    \includegraphics{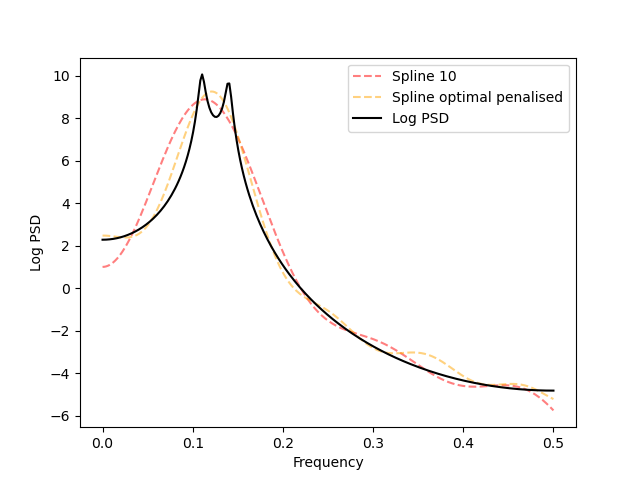}
    \caption{Analytic and estimated log PSD for 10-eigenvector and optimal decomposition of the smoothing spline model for the Percival-Walden AR(4), defined in Appendix \ref{appendix:PW}. The optimal and penalised optimal smoothing splines coincide, with 23 eigenvectors. One can see that the spectral estimates each only have one spectral peak, while the analytic PSD has two, providing an example where the proximity matching criterion of Section \ref{sec:multiplepeaks} fails.}
    \label{fig:PW}
\end{figure*}

In this brief section, we apply our methodology to a rather challenging autoregressive process that has been highlighted several times in the literature \cite{Boxbook,Percival1993} and is commonly known as the Percival-Walden AR(4). This process is defined as $x_{t} = 2.7607 x_{t-1} - 3.8106 x_{t-2} + 2.6535 x_{t-3} -0.9238 x_{t-4} + \epsilon_{t}$, simulated over length $n=1024.$ As in Section \ref{simulation}, we simulate the process, and validate our spectral density estimates against the known analytic power spectrum. In this experiment, the optimal and penalised optimal smoothing spline coincide, with 23 eigenvectors. The spectral estimates are plotted in Figure \ref{fig:PW} while validation metrics are provided in Table \ref{tab:PWmetrics}.

This experiment also provides an example where the proximity matching criterion of Section \ref{sec:multiplepeaks} fails. We can simply observe that the analytic log power spectrum $\mathbf{g}$ has two peaks, while the spectral estimate $\hat{\mathbf{g}}$ for both Spline$_{10}$ and the (penalised) optimal smoothing spline have one peak each. As such, Table \ref{tab:PWmetrics} includes the values of the semi-metric presented in (\ref{eq:MJ}) in Section \ref{sec:MJdefinition}.  We observe that the (penalised) optimal smoothing spline provides a better approximation of the amplitudes of the two peaks than the existing method of Rosen et al.

\begin{table*}
\centering
\begin{tabular}{ |p{1.9cm}||p{1cm}|p{.8cm}|p{1.8cm}|p{3.5cm}|p{2.5cm}|p{2cm}|p{2cm}|  }
 \hline
 \multicolumn{8}{|c|}{Percival-Walden process: validation metrics} \\
 \hline
 Method & RMSE & MAE & Wasserstein & $|\argmax g - \argmax \hat{g}|$ & $|\max g - \max \hat{g} |$ & Frequency distance (\ref{eq:MJ}) & Amplitude distance (\ref{eq:MJ})  \\
 \hline
 Spline$_{10}$ & 0.91    & 0.61 & 0.51 & 0.002 & 1.17 & 0.0083  & 0.8575 \\
 Spline$_\text{optimal}$ & 0.50   & 0.36 & 0.31 & 0.01 & 0.81 & 0.013 & 0.136\\
 \hline	
\end{tabular}
\caption{Results for synthetic experiments on the Percival-Walden AR(4), defined in Appendix \ref{appendix:PW}. The optimal and penalised optimal splines coincide, with 23 eigenvectors. In this instance, the proximity matching criterion from Section \ref{sec:multiplepeaks} fails, so the distance between sets of peaks in Eq. (\ref{eq:MJ}) must be used. We include this distance both between sets of frequencies and amplitudes.}
\label{tab:PWmetrics}
\end{table*}


\section{Discussion of select existing methods}
Alongside the statistics community, many signal processing practitioners and engineers have long been interested in the study of time series' power spectra. Thus, it is worth noting the differences between a framework such as ours and more traditional, signal processing-based methods for power spectral density estimation.

We begin by describing Welch's method, which is based upon Barlett's method. Welch's method aims to reduce noise in the resulting power spectral density estimate by sacrificing the degree of frequency resolution. The data is sub-divided into overlapping segments, in each of which, a modified periodogram is computed. Each modified periodogram is averaged to produce a final estimate of the power spectral density. There are two model parameters in Welch's method: the length of each segment and the degree of overlap in data points between adjacent segments.

Practically, Welch's method has several limitations in comparison to Bayesian methods. First, the estimated  power spectral density when using methods such as these may be less smooth. For scientists hoping to make observations with respect to the maximum amplitude and corresponding frequencies for any underlying time series, the often rough nature of Welch's method may make inference more difficult. It is common in the Bayesian statistics literature to use a flexible prior function on the log of the power spectrum, such as a Gaussian process. The smoothness of the Gaussian process may be highly dependent on the covariance structure chosen by the modeller. Applying covariance functions such as the squared exponential and select Matern family variants allows for smooth interpolation in the resulting power spectral density estimate. Furthermore, recent research has shown that the variance is not a monotonic decreasing function of the fraction of overlap within adjacent segments \cite{Barbe2010}. Second, Welch's method is unable to algorithmically partition the time series based on changes in the power spectral density. Procedures such as the RJMCMC introduced in this paper identify points in time where the power spectral density has changed. Hence, one would be unable to determine locations in the time domain which correspond to changes in the underlying periodic nature of a process, if one were to use Welch's method.

That said, many practitioners in the signal processing literature use techniques such as wavelets in the case of implementing spectral density estimation in a nonstationary setting. For instance, the continuous wavelet transform has been applied for spectral analysis in nonstationary signals. Wavelets overcome the obvious limitation with Fourier transformation-driven methods, where abrupt changes in time series' behaviour is difficult to capture (due to its underlying construction as a sum of sinusoidal waves). Unlike sine waves, which smoothly oscillate, wavelets are derived from ``step functions'' that exist for a finite duration, allowing for the efficient capture of abrupt changes in modelling tasks. Third, many would argue that a Bayesian framework such as ours provides a more principled approach to uncertainty quantification than frameworks such as Welch's method. The methodology proposed in this paper consists of uncertainty surrounding the power spectral density estimate, in addition to uncertainty surrounding the change point location. One clear advantage of Welch's method in comparison to the method we have proposed (and other MCMC-based methods) however, is a significant computational advantage. While there are certainly frequentist methods to estimate the uncertainty in traditional signal processing methods, there are always individuals who prefer the posterior distributions provided by Bayesian methods, including just philosophically.

Another commonly used framework for spectral density estimation is the multitaper method. Multitaper analysis is an extensions of traditional taper analysis, where time series are tapered before applying a Fourier transformation as a method of reducing potential bias resulting from spectral leakage. The multitaper method averages over a variety of estimators with varying window functions \cite{Thomson1982,Mann1996}. This results in a power spectrum that exhibits reduced leakage and variance, and retains important information from the initial and final sequences from the underlying time series. One major advantage of the multitaper method is that it can be applied in a fairly automatic manner, and is therefore appropriate in situations where many individual time series must be processed and a thorough analysis of each individual time series is not feasible. One possible limitation of the multitaper method is reduced spectral resolution. The multitaper method has proven to be an effective estimator in the presence of complex spectra. For example, Percival and Walden \cite{Percival1993} highlight the estimator's effectiveness in detecting two peaks in the case of their AR(4) process described in Appendix \ref{appendix:PW}. As we saw, our methodology was unable to detect the two peaks. Of course, there are many techniques currently being used in addition to Welch's method and the multitaper method described above.

Having outlined several commonly used methods for spectral density estimation, we now outline the reversible jump sampling scheme used in this paper.

\section{Reversible jump sampling scheme}
\label{appendix:RJMCMCsampling}

We follow Rosen et al. \cite{Rosen2017,Rosen2012} in our core implementation of the reversible jump sampling scheme. We remark that our method does not improve the trans-dimensional component of the model, described by the reversible jump scheme below. A time series partition is denoted $\bs{\xi}_{m} = (\xi_{0,m},...,\xi_{m,m})$ with $m$ segments. We have a vector of \emph{amplitude parameters} $\bs{\tau}_{m}^{2} = (\tau_{1,m}^{2},...,\tau_{m,m}^{2})'$ and \emph{regression coefficients} $\bs{\beta}_{m} = (\bs{\beta'}_{1,m},...,\bs{\beta'}_{m,m})$ that we wish to estimate, for the $j$th component within a partition of $m$ segments, $j=1,...,m.$ For notational simplicity, $\bs{\beta}_{j,m}, j=1,...,m,$ is assumed to include the first entry, $\alpha_{0j,m}.$ In the proceeding sections superscripts $c$ and $p$ refer to current and proposed value in the sampling scheme.

First, we describe the \textbf{between-model moves:} 
let $\bs{\theta}_{m} = (\bs{\xi}'_{m}, \bs{\tau}^{2'}_{m}, \bs{\beta'}_{m})$ be the model parameters at some point in the sampling scheme and assume that the chain starts at $(m^c, \bs{\theta}_{m^c}^{c})$. The algorithm proposes the move to $(m^p, \bs{\theta}_{m^p}^p)$, by drawing $(m^p, \bs{\theta}_{m^p}^p)$ from the proposal distribution \\ $q(m^p, \bs{\theta}_{m^p}^p|m^c, \bs{\theta}_{m^c}^c)$. That draw is accepted with probability
\begin{equation*}
    \alpha = \text{ min } \Bigg\{1, \frac{p(m^{p}, \bs{\theta}_{m^p}^{p}|\bs{x}) q(m^c, \bs{\theta}_{m^c}^{c}|m^p, \bs{\theta}_{m^p}^{p})}
    {p(m^{c}, \bs{\theta}_{m^c}^{c}|\bs{x}) q(m^p, \bs{\theta}_{m^p}^{p}|m^c, \bs{\theta}_{m^c}^{c})} \Bigg\},   
\end{equation*}
with $p(\cdot)$ referring to a target distribution, the product of the likelihood and the prior. The target and proposal distributions will vary based on the type of move taken in the sampling scheme. First, $q(m^p, \bs{\theta}_{m^{p}}^{p}| m^c, \bs{\theta}_{m^{c}}^{c})$ is described as follows:
\begin{align*}
    q(m^p, \bs{\theta}_{m^p}^{p}|m^c, \bs{\theta}_{m^{c}}^{c}) = q(m^p|m^c)  q(\bs{\theta}_{m^p}^{p}| m^p, m^c, \bs{\theta}_{m^c}^{c}) \\
    = q(m^p|m^c)  q(\bs{\xi^p_{m^p}}, \bs{\tau}_{m^p}^{2p}, \bs{\beta}_{m^p}^{p} | m^p, m^c, \bs{\theta}_{m^c}^c) \\
    = q(m^p|m^c)  q(\bs{\xi}_{m^p}^p|m^p, m^c, \bs{\theta}_{m^c}^c)  q(\bs{\tau}_{m^p}^{2p}|\bs{\xi}_{m^p}^p, m^p, m^c, \bs{\theta}_{m^c}^c) \\
    \times q(\bs{\beta}_{m^p}^{p}|\bs{\tau}_{m^p}^{2p}, \bs{\xi}_{m^p}^p, m^p, m^c, \bs{\theta}_{m^c}^c).
\end{align*}
To draw $(m^p, \bs{\theta}_{m^p}^p)$ one must first draw $m^p$, followed by $\bs{\xi}_{m^p}^p$, $\tau_{m^p}^{2p}, \text{ and } \bs{\beta}_{m^p}^{p}$. 
First, the number of segments $m^p$ is drawn from the proposal distribution $q(m^p|m^c)$. Let $M$ be the maximum number of segments and $m^{c}_{2,\text{min}}$ be the number of current segments whose cardinality is at least $2  t_{min}$ data points. The proposal is as follows:

\begin{align*}
    q(m^p = k | m^c)= \\ \left\{
                \begin{array}{ll}
                  1/2 \text{ if } k = m^c - 1, m^c + 1 \text{ and } m^{c} \neq 1, M, m_{2,\text{min}}^c \neq 0\\
                  1 \text{ if } k = m^{c}-1 \text{ and } m^{c} = M \text{ or } m_{2,\text{min}}^{c} = 0 \\
                  1 \text{ if } k = m^{c} + 1 \text{ and } m^{c} = 1
                \end{array}
              \right.
\end{align*}
Conditional on the proposed model $m^p$, a new partition $\bs{\xi}_{m^p}^{p}$, a new vector of covariance amplitude parameters $\bs{\tau}_{m^p}^{2p}$, and a new vector of regression coefficients, $\bs{\beta}_{m^p}^{p}$ are proposed. In \cite{Rosen2012}, $\tau^{2}$ is referred to as a smoothing parameter. To impact the smoothness of the covariance function, the parameter would have to impact pairwise operations. Given that $\tau^{2}$ sits outside the covariance matrix, we will refer to $\tau^{2}$ as an amplitude parameter (akin to signal variance within the Gaussian process framework \cite{Rasmussen2006}). 

Now, we describe the process of the \textbf{birth} of new segments. Suppose that $m^p = m^c + 1$. A time series partition,
    \begin{align*}
    \bs{\xi}^{p}_{m^p} = (\xi^c_{0,m^c},...,\xi^c_{k^{*}-1,m^c},\xi_{k^{*},m^{p}}^{p}, \xi_{k^{*},m^{c}}^{c},...,\xi_{m^c,m^c}^{c})    
    \end{align*}
    is drawn from the proposal distribution $q(\bs{\xi}_{m^p}^{p}|m^p, m^c, \bs{\theta}_{m^c}^c)$. The algorithm proposes a partition by first selecting a random segment $j = k^{*}$ to split. Then, a point $t^{*}$ within the segment $j=k^{*}$ is randomly selected to be the proposed partition point. This is subject to the constraint, \\ $\xi_{k^{*}-1, m^c}^{c} + t_{\text{min}} \leq t* \leq \xi_{k^{*},m^c}^c - t_{\text{min}}$. The proposal distribution is computed as follows:
    \begin{align*}
        q(\xi_{j,m^p}^{p} = t^{*} | m^p, m^c, \bs{\xi}_{m^c}^{c}) =  p(j=k^{*} | m^p, m^c, \bs{\xi}_{m^c}^{c}) \\
         p(\xi_{k^{*}, m^p}^{p} = t^{*} | j=k^{*}, m^p, m^c, \bs{\xi}_{m^c}^c) \\
        =  \frac{1}{m_{2 \text{min}}^{c}(n_{k^{*}, m^c} - 2t_{\text{min}}+1)}.
    \end{align*}
The vector of amplitude parameters
    \begin{align*}
        \tau_{m^p}^{2p} = (\tau_{1,m^c}^{2c},...,\tau_{k^{*}-1,m^c}^{2c}, \\ \tau_{k^{*},m^p}^{2p}, \tau_{k^{*}+1,m^p}^{2p}, \tau_{k^{*}+1,m^c}^{2c},...,\tau_{m^c, m^c}^{2c})
    \end{align*}
    is drawn from the proposal distribution \\ $q(\bs{\tau}_{m^p}^{2p}|m^p, \bs{\xi}_{m^p}^p, m^c, \bs{\theta}_{m^c}^c) = q(\bs{\tau}_{m^p}^{2p}|m^p, \bs{\tau}_{m^c}^{2c}).$ The algorithm is based on the reversible jump algorithm of Green \cite{GREEN1995}. It draws from a uniform distribution $u \sim U[0,1]$ and defines $\tau_{k^{*}, m^p}^{2p}$ and $\tau_{k^{*}+1, m^p}^{2p}$ in terms of $u$ and $\tau_{k^{*}, m^c}^{2c}$ as follows:
    \begin{align}
    \label{eq:birth1}
        \tau_{k^{*, m^p}}^{2p} =   \frac{u}{1-u}\tau_{k^{*}, m^c}^{2c}; \\
        \tau_{k^{*}+1, m^p}^{2p} =   \frac{1-u}{u}\tau_{k^{*}, m^c}^{2c}.
        \label{eq:birth2}
    \end{align}
The vector of coefficients
    \begin{align*}
        \bs{\beta}_{m^p}^p = (\bs{\beta}_{1,m^c}^{c},...,\bs{\beta}_{k^{*}-1,m^c}^{c}, \\  \bs{\beta}_{k^{*}, m^p}^p, \bs{\beta}_{k^{*}+1,m^p}^{p}, \bs{\beta}_{k^{*}+1,m^c}^{c},...,\bs{\beta}_{m^c,m^c}^{c})
    \end{align*}
    is drawn from the proposal distribution \\ $q(\bs{\beta}_{m^p}^p|\bs{\tau}_{m^p}^{2p},\bs{\xi}_{m^p}^{2p},m^p, m^c, \bs{\theta}_{m^c}^c) = q(\bs{\beta}_{m^p}^{p}|\bs{\tau}_{m^p}^{2p}, \bs{\xi}_{m^p}^p, m^p)$. The pair of vectors $\bs{\beta}_{k^{*}, m^p}^p$ and $\bs{\beta}_{k^{*}+1, m^p}^p$ are drawn from Gaussian approximations to the respective posterior conditional distributions $p(\bs{\beta}_{k^{*}, m^p}^p|\bs{x}_{k^{*}}^p, \tau_{k^{*}, m^p}^{2p}, m^p)$ and \\ $p(\bs{\beta}_{k^{*}+1, m^p}^{p}|\bs{x}_{k^{*}+1}^p, \tau_{k^{*}+1, m^p}^{2p}, m^p)$, respectively. Here, $\bs{x}_{k^{*}}^p$ and $\bs{x}_{k^{*}+1}^p$ refer to the subsets of the time series with respective segments $k^{*}$ and $k^{*}+1$. $\bs{\xi}_{m^p}^p$ will determine $\bs{x_{*}}^p = (\bs{x}_{k^{*}}^{p'}, \bs{x}_{k^{*}+1}^{p'})'$. For the sake of exposition, we provide the following example: the coefficient $\bs{\beta}_{k^{*}, m^p}^{p}$ is drawn from the Gaussian distribution $N(\bs{\beta}_{k^{*}}^{\text{max}}, \Sigma_{k^{*}}^{\text{max}})$, where $\bs{\beta}_{k^{*}}^{\text{max}}$ is defined as 
    \begin{align*}
         \argmax_{\bs{\beta}_{k^{*}, m^p}^{p}} p(\bs{\beta}^p_{k^{*}, m^p}|\bs{x}_{k^{*}}^p, \tau_{k^{*}, m^p}^{2p}, m^p)
    \end{align*} and 
    \begin{align*}
       \Sigma_{k^{*}}^{\text{max}} =-\Bigg \{ \pdv{\log p(\bs{\beta}_{k^{*}, m^p}^p | \bs{x}_{k^{*}}^p, \tau_{k^{*}, m^p}^{2p}, m^p)}{\bs{\beta}_{k^{*}, m^p}^{p}}{\bs{\beta}_{k^{*}, m^p}^{p'}} \Bigg|_{\bs{\beta}_{k^{*}, m^p}^{p} = \bs{\beta}_{k^{*}}^{\text{max}}}  \Bigg \}_.^{-1}
    \end{align*}
    For the birth move, the probability of acceptance is $\alpha = \min\{1,A\}$, where $A$ is equal to
    \begin{align*}
        \Bigg| \pdv{(\tau_{k^{*}, m^p}^{2p}, \tau_{k^{*}+1, m^p}^{2p})}{(\tau_{k^{*}, m^c, u}^{2c})}  \Bigg|\frac{p(\bs{\theta}_{m^p}^p|\bs{x}, m^p) p(\bs{\theta}_{m^p}^p|m^p)p(m^p)}{p(\bs{\theta}_{m^p}^p|\bs{x}, m^p) p(\bs{\theta}_{m^c}^c|m^c)p(m^c)} \\
        \times \frac{p(m^{c}|m^p)p(\bs{\beta}_{k^{*}, m^c}^{c})}{p(m^p|m^c)p(\xi_{k^{*}, m^p}^{m^p}|m^p, m^c) p(u) p(\bs{\beta}^p_{k^{*}, m^p})p(\bs{\beta}_{k^{*}+1,m^{p}}^{p})}.  
    \end{align*}
    Above, $p(u) = 1, 0 \leq u \leq 1,$ while $p(\bs{\beta}_{k^{*}, m^p}^{p})$ and $p(\bs{\beta}_{k^{*}+1, m^p}^{p})$ are Gaussian proposal distributions $N(\bs{\beta}_{k^{*}}^{\text{max}}, \Sigma_{k^{*}}^{\text{max}})$ and \\ $N(\bs{\beta}_{k^{*}+1}^{\text{max}}, \Sigma_{k^{*}+1}^{\text{max}})$, respectively. The Jacobian is computed as
    \begin{align*}
        \bigg| \pdv{(\tau_{k^{*}, m^p}^{2p}, \tau_{k^{*}+1, m^p}^{2p})}{(\tau^{2c}_{k^{*}, m^c}, u)} \bigg| = \frac{2 \tau_{k^{*}m^c}^{2c}}{u(1-u)} = 2(\tau_{k^{*}, m^p}^{p} + \tau_{k^{*}+1, m^p}^{p})^{2}.
    \end{align*}

Next, we describe the process of the \textbf{death} of new segments, that is, the reverse of a birth move, where  $m^p = m^c - 1$. A time series partition
\begin{align*}
\bs{\xi}_{m^p}^{p} = (\xi_{0,m^c}^{c},...,\xi_{k^{*}-1,m^c}^{c}, \xi_{k^{*}+1,m^c}^{c},...,\xi_{m^c,m^c}^{c}),
\end{align*}
is proposed by randomly selecting a single partition from $m^c - 1$ candidates, and removing it. The partition point selected for removal is denoted $j=k^{*}$. There are $m^c -1$ possible segments available for removal among the $m^c$ segments currently in existence. The proposal may choose each partition point with equal probability, that is,
\begin{align*}
    q(\xi_{j, m^p}^p|m^p, m^c, \bs{\xi}_{m^c}^c) = \frac{1}{m^c - 1}.
\end{align*}
The vector of amplitude parameters 
\begin{align*}
\bs{\tau}_{m^p}^{2p} = (\tau_{1, m^c}^{2c},...,\tau_{k^{*}-1,m^c}^{2c},\tau_{k^{*},m^p}^{2c}, \tau_{k^{*}+2,m^c}^{2c},...,\tau_{m^c,m^c}^{2c})    
\end{align*}
 is drawn from the proposal distribution \\ $q(\bs{\tau}_{m^p}^{2p}|m^p, \bs{\xi}_{m^p}^p, m^c, \bs{\theta}_{m^c}^{c}) = q(\bs{\tau}_{m^p}^{2p}| m^p, \bs{\tau}_{m^c}^{2c})$. One amplitude parameter $\tau_{k^{*}, m^p}^{2p}$ is formed from two candidate amplitude parameters, $\tau_{k^{*},m^c}^{2c}$ and $\tau_{k^{*}+1,m^c}^{2c}$. This is done by reversing the equations \ref{eq:birth1} and \ref{eq:birth2}. That is,
\begin{align*}
    \tau_{k^{*}, m^p}^{2p} = \sqrt{\tau_{k^{*}, m^{c}}^{2c} \tau_{k^{*}+1, m^c}^{2c}}.
\end{align*}
Finally, the vector of regression coefficients,
\begin{align*}
    \bs{\beta}_{m^p}^{p} = (\beta_{1,m^c}^{c},...,\beta_{k^{*}-1,m^c}^{c}, \beta_{k^{*}, m^p}^{p}, \beta_{k^{*}+2,m^c}^{c},...,\beta_{m^c,m^c}^{c})
\end{align*}
is drawn from the proposal distribution \\  $q(\bs{\beta}_{m^p}^p|\bs{\tau}_{m^p}^{2p}, \bs{\xi}_{m^p}^p, m^p, m^c, \theta_{m^c}^c) = q(\bs{\beta}_{m^p}^{p}|\bs{\tau}_{m^p}^{2p}, \bs{\xi}_{m^p}^p, m^p)$. The vector of regression coefficients is drawn from a Gaussian approximation to the posterior distribution \\ $p(\beta_{k^{*},m^p}|\bs{x}, \tau_{k^{*}, m^p}^{2p}, \bs{\xi}^p_{m^p}, m^p)$ following the same procedure for the vector of coefficients in the birth step. The probability of acceptance is the inverse of the analogous birth step. If the move is accepted then the following updates occur: $m^c=m^p$ and $\bs{\theta}_{m^c}^c = \bs{\theta}_{m^p}^{p}$.

Finally, we describe the \textbf{within-model moves:}
henceforth, $m$ is fixed; accordingly, notation describing the dependence on the number of segments is removed. There are two parts to a within-model move. First, a segment relocation is performed, and conditional on the relocation, the basis function coefficients are updated. The steps are jointly accepted or rejected with a Metropolis-Hastings step. The amplitude parameters are updated within a separate Gibbs sampling step. 

The chain is assumed to be located at $\bs{\theta}^{c} = (\bs{\xi}^{c}, \bs{\beta}^{c})$. The proposed move $\bs{\theta}^p = (\bs{\xi}^p, \bs{\beta}^p)$ is as follows: first, a partition point $\xi_{k^{*}}$ is selected for relocation from $m-1$ candidate partition points. Next, a position within the interval $[\xi_{k^{*}-1}, \xi_{k^{*}+1}]$ is selected, subject to the fact that the new location is at least $t_{\text{min}}$ data points away from $\xi_{k^{*}-1}$ and $\xi_{k^{*}+1}$, so that
\begin{align*}
    \Pr(\xi^p_{k^{*}}=t) = \Pr (j=k^{*})  \Pr (\xi_{k^{*}}^{p}=t|j=k^{*}), 
\end{align*}
where $\Pr(j=k^{*}) = (m-1)^{-1}$. A mixture distribution for $\Pr(\xi_{k^{*}}^p=t|j=k^{*})$ is constructed to explore the space most efficiently, so
\begin{align*}
    \Pr(\xi_{k^{*}}^{p}=t|j=k^{*}) = \\ \pi q_1 (\xi_{k^{*}}^p = t| \xi_{k^{*}}^{c}) + (1-\pi) q_2 (\xi_{k^{*}}^p=t|\xi_{k^{*}}^c),
\end{align*}
where $q_1(\xi_{k^{*}}^p = t| \xi_{k^{*}}^c) = (n_{k^{*}} + n_{k^{*}+1}-2t_{\text{min}} + 1)^{-1}, \xi_{k^{*}-1} + t_{\text{min}} \leq t \leq \xi_{k^{*}+1} - t_{\text{min}}$ and 

\begin{align*}
    q_2(\xi_{k^{*}}^p = t|\xi_{k^{*}}^{c})= \\
\left\{
            \begin{array}{ll}
              0 \text{ if } |t-\xi^c_{k^{*}}| > 1  \\
            1/3 \text{ if } |t-\xi^c_{k^{*}}| \leq 1, n_{k^{*}} \neq t_{\text{min}} \text{ and } n_{k^{*}+1} \neq t_{\text{min}}  \\ 
            1/2 \text{ if } t-\xi_{k^{*}}^{c} \leq 1, n_{k^{*}} = t_{\text{min}} \text{ and } n_{k^{*}+1} \neq t_{\text{min}} \\
            1/2 \text{ if } \xi_{k^{*}}^{c} - t \leq 1, n_{k^{*}} \neq t_{\text{min}} \text{ and } n_{k^{*}+1} = t_{\text{min}} \\
            1 \text{ if } t = \xi_{k^{*}}^{c}, n_{k^{*}} = t_{\text{min}} \text{ and } n_{k^{*}+1} = t_{\text{min}}
            \end{array}
          \right.
\end{align*}
The support of $q_1$ has $n_{k^{*}} + n_{k^{*}+1} - 2t_{\text{min}} + 1$ data points while $q_2$ has at most three. The term $q_2$ alone would result in a high acceptance rate for the Metropolis-Hastings, but it would explore the parameter space slowly. The $q_1$ component allows for larger jumps, and produces a compromise between a high acceptance rate and thorough exploration of the parameter space. 

 Next, $\bs{\beta^{p}_{j}}, j=k^{*}, k^{*}+1$ is drawn from an approximation to $\prod^{k^{*}+1}_{j=k^{*}} p(\bs{\beta}_j|\bs{x}_j^p, \tau_j^{2})$, following the analogous step in the between-model move. The proposal distribution, which is evaluated at $\bs{\beta}^{p}_j, j=k^{*}, k^{*}=1$, is 
\begin{align*}
    q(\bs{\beta}_{*}^{p}|\bs{x}_{*}^p, \bs{\tau}_{*}^{2}) = \prod^{k^{*}+1}_{j=k^{*}} q(\bs{\beta}_{j}^p|\bs{x}_j^p, \tau_j^{2}),
\end{align*}
where $\bs{\beta}_{*}^p = (\bs{\beta}^{p'}_{k^{*}}, \bs{\beta}^{p'}_{k^{*}+1})'$ and $\bs{\tau}_{*}^{2} = (\tau^{2}_{k^{*}}, \tau^{2}_{k^{*}+1})'$. The proposal distribution is evaluated at current values of $\bs{\beta}_{*}^{c} = (\beta^{c'}_{k^{*}}, \beta^{c'}_{k^{*}+1})'$. $\beta_{*}^p$ is accepted with probability
\begin{equation*}
    \alpha = \min \Bigg\{ 1, \frac{p(\bs{x}_{*}^p|\bs{\beta}_{*}^{p}) p(\bs{\beta}_{*}^{p}|\bs{\tau}_{*}^{2}) q(\bs{\beta}_{*}^{c}|\bs{x}^{c}_{*}, \bs{\tau}_{*}^{2})} {p(\bs{x}_{*}^c|\bs{\beta}_{*}^{c}) p(\bs{\beta}_{*}^{c}|\bs{\tau}_{*}^{2}) q(\bs{\beta}_{*}^{p}|\bs{x}^{p}_{*}, \bs{\tau}_{*}^{2})} \Bigg\},
\end{equation*}
where $\bs{x}_{*}^{c} = (\bs{x}^{c'}_{k^{*}},\bs{x}^{c'}_{k^{*}+1})$. When the draw is accepted, update the partition and regression coefficients $(\xi^{c}_{k^{*}}, \beta_{*}^{c}) = (\xi^{p}_{k^{*}}, \beta_{*}^{p})$.
Finally, draw $\tau^{2p}$ from 
\begin{align*}
    p(\tau_{*}^{2}|\bs{\beta}_{*}) = \prod^{k^{*}+1}_{j=k^{*}} p(\tau_j^{2}|\beta_j).
\end{align*}
This is a Gibbs sampling step, and accordingly the draw is accepted with probability 1.

\section{Metropolis-Hastings algorithm}
\label{appendix:MH}

In this section, we describe the Metropolis-Hastings algorithm used in the stationary case for our simulation study (Section \ref{simulation}). As seen in Appendix \ref{appendix:RJMCMCsampling}, the above RJMCMC reduces to a Metropolis-Hastings in the absence of the between-model moves.

We estimate the log of the spectral density by its posterior mean via a Bayesian approach and an adaptive MCMC algorithm:
\begin{equation}
\E (\mathbf{g}|\mathbf{y}) = \int \E (\mathbf{g}|\mathbf{y}, \mathbf{\theta}) p(\mathbf{\theta}|\mathbf{y}) d\theta \,\, \simeq \,\, \frac{1}{M} \sum_{j=1}^{M} \E (\hat{\mathbf{g}}|\mathbf{y}, \mathbf{\theta}^{j}).
\end{equation}
Here, $M$ is the number of post burn-in iterations in the MCMC scheme;  $\mathbf{\theta}^{j}$ are samples taken from the posterior distribution $p(\mathbf{\theta}|\mathbf{y})$; $p(\mathbf{\theta})$ is taken from a Gaussian distribution $N(\mu, \sigma^{2})$ centred around $\mu=\theta^{[c]}$ that maximises the log marginal likelihood; $\sigma$ is chosen arbitrarily; and $\hat{\mathbf{g}}$ is the forecast log spectrum.

 Monte Carlo algorithms have been highly prominent for estimation of hyperparameters and spectral density in a nonparametric Bayesian framework. Metropolis \cite{Metropolis1953} first proposed the Metropolis algorithm; this was generalised by Hastings in a more focused, statistical context \cite{Hastings1970}.
The random walk Metropolis-Hastings algorithm aims to sample from a target density $\pi$, given some candidate transition probability function $q(x,y)$. In our context, $\pi$ represents the Whittle likelihood function multiplied by respective priors. The acceptance ratio is:
\begin{equation}
    \alpha (x,y) =
    \begin{cases}
      \min \bigg( \frac{\pi(y)q(y,x)}{\pi(x) q(x,y)}, 1 \bigg) \text{ if } \pi(x) q(x,y) > 0 \\
      1  \text{ if } \pi(x) q(x,y) = 0.
    \end{cases}
\end{equation}

Our MCMC scheme calculates the acceptance ratio every 50 iterations; 
based on an ideal acceptance ratio \cite{Roberts2004}, the step size is adjusted for each hyperparameter. The power spectrum and hyperparameters for the GP covariance function are calculated in each iteration of our sampling scheme and integrated over. 

First, we initialise the values of our GP hyperparameters, $\theta^{[c]}$, our log PSD $\hat{g}^{c}$, our random perturbation size $s^{2}$, and the adaptive adjustment for our step size $\xi$. Starting values for our GP hyperparameters are chosen based on maximising the marginal Whittle likelihood,
\begin{equation}
\theta^{[c]} = \argmax_{\theta} - (2 \pi)^{-m/2} \prod_{j=0}^{m-1} \exp\left({-\frac{1}{2}\left[\log f(\nu_{j}) + \frac{I(\nu_{j})}{f(\nu_{j})}\right]}\right).
\end{equation}

The latent log PSD is modelled with a zero mean GP, That is, $\mathbf{g} \sim GP(0,k_{\theta}(x,x'))$, and we follow the notation of \cite{Rasmussen2006} where $k(x,x')$ refers to any respective kernel. The adaptive MCMC algorithm samples from the posterior distribution of the log PSD and the posterior distribution of any candidate covariance function's hyperparameters. First, the current and proposed values for the mean and covariance of the Gaussian process are computed. That is, the mean is computed:

\begin{equation}
\hat{g^{c}} = k_{\theta^c}(x',x)[k_{\theta^c}(x,x) + \sigma^{2}I]^{-1}\log I(\mathbf{\nu}),
\end{equation}
and the covariance is computed:
\begin{equation}
\hat{V^{c}} = k_{\theta^c}(x',x')-k_{\theta^c}(x',x)[k_{\theta^c}(x,x) + \sigma^{2}I]^{-1}k_{\theta^c}(x,x').
\end{equation}
New proposals for GP hyperparameters are determined via a random walk proposal distribution. A zero-mean Gaussian distribution is used to generate candidate perturbations, where $s^2$ is the variance of this Gaussian. That is 
\begin{equation}
\theta^{p} \xleftarrow[]{} q(\theta^p|\theta^c).
\end{equation}

Having drawn the proposed GP hyperparameters, a proposed mean and covariance function of the log PSD are drawn from the posterior distribution of the GP. Both the proposed mean and covariance are computed similarly to the current values, simply replacing the values of the hyperparameters $\theta^{c} \xleftarrow[]{} \theta^{p}$. So, the proposed mean of the log PSD is, 
\begin{equation}
\hat{g^{p}} = k_{\theta^p}(x',x)[k_{\theta^p}(x,x) + \sigma^{2}I]^{-1}\log I(\mathbf{\nu})
\end{equation}
and the proposed covariance is computed as follows
\begin{equation}
\hat{V^{p}} = k_{\theta^p}(x',x')-k_{\theta^p}(x',x)[k_{\theta^p}(x,x) + \sigma^{2}I]^{-1}k_{\theta^p}(x,x').    
\end{equation}
Having computed the proposed and current values of the log PSD, we update the current log PSD based on the Metropolis-Hastings transition kernel. First, we sample  from a uniform distribution $u \sim U(0,1)$ and compute our acceptance ratio,
\begin{equation}
\alpha = \min \left(1, \frac{p(\log I(\mathbf{\nu}) \mid \theta^{p}, \hat{g}^{p}) p(\hat{g}^{p}) q(\hat{g}^{c} \mid \log I(\mathbf{\nu}), \theta^{p}))}{p(\log I(\mathbf{\nu}) \mid \theta^{c}, \hat{g}^{c}) p(\hat{g}^{c}) q(\hat{g}^{p} \mid \log I(\mathbf{\nu}), \theta^{c}))}\right).
\end{equation}

$p(\log I(\mathbf{\nu}) \mid \theta^{p}, \hat{g}^{p})$ is our Whittle likelihood computation, the probability of the log PSD conditional on hyperparameters $\theta$ and our candidate estimate of the latent log PSD $\hat{g}$. $p(\hat{g}^{p})$ represents the prior distribution on our latent log PSD and $q(\hat{g}^{c} \mid \log I(\mathbf{\nu}), \theta^{p}))$ is our proposal distribution, representing the probability of the estimated log PSD conditional on the log periodogram and GP hyperparameters.

Should $u < \alpha$, we update the current values of the log PSD mean and spectrum to the proposed values. That is, 
\begin{align}
\hat{g}^{c+1} \xleftarrow[]{} \hat{g}^{p} \\
\hat{V}^{c+1} \xleftarrow[]{} \hat{V}^{p}.    
\end{align}

If $u > \alpha$, both the mean and variance of the log PSD are kept at their current values, 
 
 \begin{align}
 \hat{g}^{c+1} \xleftarrow[]{} \hat{g}^{c} \\    
 \hat{V}^{c+1} \xleftarrow[]{} \hat{V}^{c}. 
 \end{align}

Importantly, modelling the log PSD with a GP prior does not mean that we are assuming a Gaussian error distribution around the spectrum. In actuality, proposed spectra are accepted and rejected through a Metropolis-Hastings procedure - resulting in log PSD samples being drawn from the true posterior distribution of the log PSD, $\log(\exp (1))$. Having sampled the log PSD, we then accept/reject candidate GP hyperparameters with another Metropolis-Hastings step. Our acceptance ratio is, 
\begin{equation}
\alpha = \min \left(1, \frac{p(\log I(\mathbf{\nu}) \mid \theta^{p}) p(\theta^{c} \mid \theta^{p})}{p(\log I(\mathbf{\nu}) \mid \theta^{c}) p(\theta^{p} \mid \theta^{c})}\right),    
\end{equation}
where $p(\log I(\mathbf{\nu}) \mid \theta)$ represents the Whittle likelihood modelling the probability of the log PSD, $\log I(\mathbf{\nu})$, conditional on hyperparameters $\theta$. $p(\theta^{p} \mid \theta)$ is the prior distribution we place over GP hyperparameters, $\theta$. Note that in this particular case, the symmetric proposal distributions cancel out and our algorithm reduces simply to a Metropolis ratio. Again we follow the standard Metropolis-Hastings acceptance decision. If $u < \alpha$, 
\begin{equation}
\theta^{c+1} \xleftarrow[]{} \theta^{p},   
\end{equation}
and the current hyperparameter values assume proposed values. Alternatively, if $u > \alpha$, 
\begin{align}
\theta^{c+1} \xleftarrow[]{} \theta^{c},   
\end{align}
current value of the hyperparameters are not updated. Finally, following \cite{Roberts2004} we implement an adaptive step-size within our random walk proposal. Every 50 iterations within our simulation, we compute the trailing acceptance ratio. An optimal acceptance ratio, $\text{Acc}^{Opt}$ of 0.234 is targeted. If the acceptance ratio is too low, indicating that the step size may be too large, then the step size is systematically reduced. If $\text{Acceptance Ratio}_{(j-49):j} \forall j \in \{50,100,150,...,10000\} < \text{Acc}^{Opt}$,
\begin{equation}
s^{2} \xleftarrow[]{} s^{2} - \xi.
\end{equation}
If the acceptance ratio is too high, indicating that the step size may be too small then step size is systematically increased. That is, when $\text{Acceptance Ratio}_{(j-49):j} \forall j \in  \{50,100,...,10000\} < \text{Acc}^{Opt}$,
\begin{equation}
s^{2} \xleftarrow[]{} s^{2} + \xi.
\end{equation}
Finally, the log PSD and the respective analytic uncertainty bounds are determined by computing the median of the samples generated from the sampling procedure,
\begin{align}
\mathcal{U}^{\text{final}}_{0.025} = \text{median}(\mathcal{U}^{5000:10 000}_{0.025}) \\
\hat{g} = \text{median}(\hat{g}^{5000:10 000}) \\
\mathcal{U}^{\text{final}}_{0.975} = \text{median}(\mathcal{U}^{5000:10 000}_{0.975}). 
\end{align}

\section{Turning point algorithm}
\label{appendix:TPA}


In this section, we provide more details for the identification of non-trivial peaks (local maxima). We aim to outline a broad and flexible framework for this purpose, in which the exact procedure may be altered according to the specific application. For example, one way to determine peaks of a given spectral estimate is simply by inspection. We aim to provide an algorithmic framework as an alternative to this.

Let $\mathbf{g}$ be an analytic or estimated log power spectral density function. We may begin, if necessary, by applying additional smoothing to this function (though this step is optional and can be omitted). Following \cite{james2021_CovidIndia}, we apply a two-step algorithm to the (possibly smoothed) function $\mathbf{g}$, defined on $\nu_j=\frac{j}{n}, j=0,1,...,m-1$. The first step produces an alternating sequence of local minima (troughs) and local maxima (peaks), which may include some immaterial turning points. The second step refines this sequence according to chosen conditions and parameters. The most important conditions to initially identify a peak or trough, respectively, are the following:
\begin{align}
\label{baddefnpeak}
g(\nu_{j_0})&=\max\{g(\nu_j): \max(0,j_0 - l) \leq j \leq \min(j_0 + l,m-1)\},\\
\label{baddefntrough}g(\nu_{j_0})&=\min\{g(\nu_j): \max(0,j_0 - l) \leq t \leq \min(j_0 + l,m-1)\},
\end{align}
where $l$ is a parameter to be chosen. Defining peaks and troughs according to this definition alone has some flaws, such as the potential for two consecutive peaks.

Instead, we implement an inductive procedure to choose an alternating sequence of peaks and troughs. Suppose $j_0$ is the last determined peak. We search in the period $j>j_0$ for the first of two cases: if we find a time $j_1>j_0$ that satisfies (\ref{baddefntrough}) as well as a non-triviality condition $g(j_1)<g(j_0)$, we add $j_1$ to the set of troughs and proceed from there. If we find a time $j_1>j_0$ that satisfies (\ref{baddefnpeak}) and  $g(t_0)\geq g(j_1)$, we ignore this lower peak as redundant; if we find a time $j_1>j_0$ that satisfies (\ref{baddefnpeak}) and  $g(j_1) > g(j_0)$, we remove the peak $j_0$,  replace it with $j_1$ and continue from $j_1$. A similar process applies from a trough at $j_0$. 

As a side remark, for an analytic log PSD $\mathbf{g}$, we could simply use the analytical and differentiable form to find critical points as an alternative.

With either possibility, at this point, the function is assigned an alternating sequence of troughs and peaks. However, some turning points are immaterial and should be removed. Here, the framework can incorporate a flexible series of options to refine the set of peaks.

As mentioned in Section \ref{sec:multiplepeaks}, one relatively simple option is simply to remove any local maximum (peak) $\hat{\rho}$ of $\hat{\mathbf{g}}$ with $\hat{g}(\hat{\rho})<\max \hat{\mathbf{g}} - \delta$, for some sensible constant $\delta$. In our experiments, the same results are produced for any $\delta \in [2,4]$, demonstrating the robustness of this relatively simple idea. Under an affine transformation of the original time series $X'_t = aX_t + b$, the log PSD $\hat{g}$ changes by an additive constant, so this condition is unchanged when rescaling the original data.

This relatively simple condition is sufficient for our application. For the benefit of future work, we list some alternative options for an algorithmic refinement of the peaks (besides, of course, inspection as the simplest option). In previous work, we have analysed functions $\nu(t)$ that were necessarily valued only in non-negative reals. Thus, one may simply linearly the log PSD $\mathbf{g}$ so that its minimum value is zero. Then, numerous options exist for refinement of non-trivial peaks (and troughs).

For example, let $t_1<t_3$ be two peaks, necessarily separated by a trough. We select a parameter $\delta=0.2$, and if the \emph{peak ratio}, defined as $\frac{\nu(t_3)}{\nu(t_1)}<\delta$, we remove the peak $t_3$. If two consecutive troughs $t_2,t_4$ remain, we remove $t_2$ if $\nu(t_2)>\nu(t_4)$, otherwise remove $t_4$. That is, if the second peak has size less than $\delta$ of the first peak, we remove it. Alternatively, one may use this peak ratio on any peak, comparing it to the global max, rather than just comparing adjacent peaks. That is, let $t_0$ be the global maximum. Then, one could remove any peak $t_1$ with $\frac{\nu(t_1)}{\nu(t_0)} < \delta$.

Alternatively, we use appropriately defined gradient or log-gradient comparisons between points $t_1<t_2$. For example, let
\begin{align}
\label{loggrad}
  \loggrad(t_1,t_2)=\frac{\log \nu(t_2) - \log \nu(t_1)}{t_2-t_1}.
\end{align}
The numerator equals  $\log(\frac{\nu(t_2)}{\nu(t_1)})$, a "logarithmic rate of change." Unlike the standard rate of change given by $\frac{\nu(t_2)}{\nu(t_1)} -1$, the logarithmic change is symmetrically between $(-\infty,\infty)$. Let $t_1,t_2$ be adjacent turning points (one a trough, one a peak). We choose a parameter $\epsilon$;  if
\begin{align}
    |\loggrad(t_1,t_2)|<\epsilon,
\end{align}
that is, the average logarithmic change is less than $\epsilon$, we remove $t_2$ from our sets of peaks and troughs. If $t_2$ is not the final turning point, we also remove $t_1$. After these refinement steps, we are left with an alternating sequence of non-trivial peaks and troughs. Finally, for this framework, we only need the peaks, so we simply discard the troughs. 

As a final remark, only at the end is the final number $r$ of non-trivial peaks determined. It is a function not only of the log PSD function $\mathbf{g}$, but also the precise conditions used to select and refine the (non-trivial) peaks.

\bibliographystyle{spmpsci}      
\bibliography{__bibliography} 

\begin{thebibliography}{10}
\providecommand{\url}[1]{{#1}}
\providecommand{\urlprefix}{URL }
\expandafter\ifx\csname urlstyle\endcsname\relax
  \providecommand{\doi}[1]{DOI~\discretionary{}{}{}#1}\else
  \providecommand{\doi}{DOI~\discretionary{}{}{}\begingroup
  \urlstyle{rm}\Url}\fi

\bibitem{Adak1998}
Adak, S.: Time-dependent spectral analysis of nonstationary time series.
\newblock Journal of the American Statistical Association \textbf{93}(444),
  1488--1501 (1998).
\newblock \doi{10.1080/01621459.1998.10473808}

\bibitem{Barbe2010}
Barbe, K., Pintelon, R., Schoukens, J.: Welch method revisited: Nonparametric
  power spectrum estimation via circular overlap.
\newblock {IEEE} Transactions on Signal Processing \textbf{58}(2), 553--565
  (2010).
\newblock \doi{10.1109/tsp.2009.2031724}

\bibitem{Boxbook}
Box, G.E.P., Jenkins, G.M., Reinsel, G.C., Ljung, G.M.: Time Series Analysis:
  Forecasting and Control.
\newblock Wiley (2015)

\bibitem{Brockwell1991}
Brockwell, P.J., Davis, R.A.: Time Series: Theory and Methods.
\newblock Springer New York (1991).
\newblock \doi{10.1007/978-1-4419-0320-4}

\bibitem{Kohn1997}
Carter, C.K., Kohn, R.: Semiparametric {B}ayesian inference for time series
  with mixed spectra.
\newblock Journal of the Royal Statistical Society: Series B (Statistical
  Methodology) \textbf{59}(1), 255--268 (1997).
\newblock \doi{10.1111/1467-9868.00067}

\bibitem{Choudhuri2004}
Choudhuri, N., Ghosal, S., Roy, A.: Bayesian estimation of the spectral density
  of a time series.
\newblock Journal of the American Statistical Association \textbf{99}(468),
  1050--1059 (2004).
\newblock \doi{10.1198/016214504000000557}

\bibitem{Cogburn1974}
Cogburn, R., Davis, H.T.: Periodic splines and spectral estimation.
\newblock The Annals of Statistics \textbf{2}(6), 1108--1126 (1974).
\newblock \doi{10.1214/aos/1176342868}

\bibitem{Dahlhaus1997}
Dahlhaus, R.: Fitting time series models to nonstationary processes.
\newblock The Annals of Statistics \textbf{25}(1), 1--37 (1997).
\newblock \doi{10.1214/aos/1034276620}

\bibitem{Duvenaud2013}
Duvenaud, D., Lloyd, J., Grosse, R., Tenenbaum, J., Ghahramani, Z.: Structure
  discovery in nonparametric regression through compositional kernel search.
\newblock In: Proceedings of the 30th International Conference on Machine
  Learning, vol.~28, pp. 1166--1174 (2013)

\bibitem{Edwards2018}
Edwards, M.C., Meyer, R., Christensen, N.: Bayesian nonparametric spectral
  density estimation using {B}-spline priors.
\newblock Statistics and Computing \textbf{29}(1), 67--78 (2019).
\newblock \doi{10.1007/s11222-017-9796-9}

\bibitem{Eilers1996}
Eilers, P.H.C., Marx, B.D.: Flexible smoothing with {B}-splines and penalties.
\newblock Statistical Science \textbf{11}(2), 89--121 (1996).
\newblock \doi{10.1214/ss/1038425655}

\bibitem{Ganngopadhyay1999}
Gangopadhyay, A., Mallick, B., Denison, D.: Estimation of spectral density of a
  stationary time series via an asymptotic representation of the periodogram.
\newblock Journal of Statistical Planning and Inference \textbf{75}(2),
  281--290 (1999).
\newblock \doi{10.1016/s0378-3758(98)00148-7}

\bibitem{GREEN1995}
Green, P.J.: Reversible jump {M}arkov chain {M}onte {C}arlo computation and
  {B}ayesian model determination.
\newblock Biometrika \textbf{82}(4), 711--732 (1995).
\newblock \doi{10.1093/biomet/82.4.711}

\bibitem{Gu2002}
Gu, C.: Smoothing Spline {ANOVA} Models.
\newblock Springer New York (2013).
\newblock \doi{10.1007/978-1-4614-5369-7}

\bibitem{Guo2003}
Guo, W., Dai, M., Ombao, H.C., von Sachs, R.: Smoothing spline {ANOVA} for
  time-dependent spectral analysis.
\newblock Journal of the American Statistical Association \textbf{98}(463),
  643--652 (2003).
\newblock \doi{10.1198/016214503000000549}

\bibitem{HadjAmar2019}
Hadj-Amar, B., Rand, B.F., Fiecas, M., L{\'{e}}vi, F., Huckstepp, R.: Bayesian
  model search for nonstationary periodic time series.
\newblock Journal of the American Statistical Association \textbf{115}(531),
  1320--1335 (2019).
\newblock \doi{10.1080/01621459.2019.1623043}

\bibitem{Hastings1970}
Hastings, W.K.: Monte {C}arlo sampling methods using {M}arkov chains and their
  applications.
\newblock Biometrika \textbf{57}(1), 97--109 (1970).
\newblock \doi{10.1093/biomet/57.1.97}

\bibitem{james2021_MJW}
James, N., Menzies, M.: A new measure between sets of probability distributions
  with applications to erratic financial behavior.
\newblock Journal of Statistical Mechanics: Theory and Experiment
  \textbf{2021}(12), 123404 (2021).
\newblock \doi{10.1088/1742-5468/ac3d91}

\bibitem{james2021_crypto2}
James, N., Menzies, M.: Collective correlations, dynamics, and behavioural
  inconsistencies of the cryptocurrency market over time.
\newblock Nonlinear Dynamics  (2022).
\newblock \doi{10.1007/s11071-021-07166-9}

\bibitem{James2019}
James, N., Menzies, M., Azizi, L., Chan, J.: Novel semi-metrics for
  multivariate change point analysis and anomaly detection.
\newblock Physica D: Nonlinear Phenomena \textbf{412}, 132636 (2020).
\newblock \doi{10.1016/j.physd.2020.132636}

\bibitem{James2021_geodesicWasserstein}
James, N., Menzies, M., Bondell, H.: Understanding spatial propagation using
  metric geometry with application to the spread of {COVID}-19 in the {U}nited
  {S}tates.
\newblock {EPL} (Europhysics Letters) \textbf{135}(4), 48004 (2021).
\newblock \doi{10.1209/0295-5075/ac2752}

\bibitem{james2021_CovidIndia}
James, N., Menzies, M., Bondell, H.: Comparing the dynamics of {COVID}-19
  infection and mortality in the {U}nited {S}tates, {I}ndia, and {B}razil.
\newblock Physica D: Nonlinear Phenomena \textbf{432}, 133158 (2022).
\newblock \doi{10.1016/j.physd.2022.133158}

\bibitem{Lu2016}
Lu, J., Hoi, S.C., Wang, J., Zhao, P., Liu, Z.Y.: Large scale online kernel
  learning.
\newblock Journal of Machine Learning Research \textbf{17}(47), 1--43 (2016)

\bibitem{Mann1996}
Mann, M.E., Lees, J.M.: Robust estimation of background noise and signal
  detection in climatic time series.
\newblock Climatic Change \textbf{33}(3), 409--445 (1996).
\newblock \doi{10.1007/bf00142586}

\bibitem{Metropolis1953}
Metropolis, N., Rosenbluth, A.W., Rosenbluth, M.N., Teller, A.H., Teller, E.:
  Equation of state calculations by fast computing machines.
\newblock The Journal of Chemical Physics \textbf{21}(6), 1087--1092 (1953).
\newblock \doi{10.1063/1.1699114}

\bibitem{Paciorek2004}
Paciorek, C.J., Schervish, M.J.: Nonstationary covariance functions for
  {G}aussian process regression.
\newblock In: Proceedings of the 16th International Conference on Neural
  Information Processing Systems, p. 273–280. MIT Press (2003)

\bibitem{Percival1993}
Percival, D.B., Walden, A.T.: Spectral Analysis for Physical Applications.
\newblock Cambridge University Press (1993).
\newblock \doi{10.1017/cbo9780511622762}

\bibitem{Plagemann2008}
Plagemann, C., Kersting, K., Burgard, W.: Nonstationary {G}aussian {P}rocess
  regression using point estimates of local smoothness.
\newblock In: Machine Learning and Knowledge Discovery in Databases, pp.
  204--219. Springer Berlin Heidelberg (2008).
\newblock \doi{10.1007/978-3-540-87481-2_14}

\bibitem{arjun}
Prakash, A., James, N., Menzies, M., Francis, G.: Structural clustering of
  volatility regimes for dynamic trading strategies.
\newblock Applied Mathematical Finance \textbf{28}(3), 236--274 (2021).
\newblock \doi{10.1080/1350486x.2021.2007146}

\bibitem{Rasmussen2006}
Rasmussen, C.E., Williams, C.K.I.: Gaussian Processes for Machine Learning.
\newblock MIT Press (2005)

\bibitem{Roberts2004}
Roberts, G.O., Rosenthal, J.S.: General state space {M}arkov chains and {MCMC}
  algorithms.
\newblock Probability Surveys \textbf{1}(0), 20--71 (2004).
\newblock \doi{10.1214/154957804100000024}

\bibitem{Rosen2009}
Rosen, O., Stoffer, D.S., Wood, S.: Local spectral analysis via a {B}ayesian
  mixture of smoothing splines.
\newblock Journal of the American Statistical Association \textbf{104}(485),
  249--262 (2009).
\newblock \doi{10.1198/jasa.2009.0118}

\bibitem{Rosen2017}
Rosen, O., Wood, S., Stoffer, D.: BayesSpec: Bayesian Spectral Analysis
  Techniques (2017).
\newblock \urlprefix\url{https://CRAN.R-project.org/package=BayesSpec}.
\newblock {R} package version 0.5.3

\bibitem{Rosen2012}
Rosen, O., Wood, S., Stoffer, D.S.: {AdaptSPEC}: Adaptive spectral estimation
  for nonstationary time series.
\newblock Journal of the American Statistical Association \textbf{107}(500),
  1575--1589 (2012).
\newblock \doi{10.1080/01621459.2012.716340}

\bibitem{Thomson1982}
Thomson, D.: Spectrum estimation and harmonic analysis.
\newblock Proceedings of the IEEE \textbf{70}, 1055--1096 (1982)

\bibitem{ToddChem2}
Todd, J.F.: Recommendations for nomenclature and symbolism for mass
  spectroscopy.
\newblock International Journal of Mass Spectrometry and Ion Processes
  \textbf{142}(3), 209--240 (1995).
\newblock \doi{10.1016/0168-1176(95)93811-f}

\bibitem{ToddChem}
Todd, J.F.J.: Recommendations for nomenclature and symbolism for mass
  spectroscopy (including an appendix of terms used in vacuum technology).
  (recommendations 1991).
\newblock Pure and Applied Chemistry \textbf{63}(10), 1541--1566 (1991).
\newblock \doi{10.1351/pac199163101541}

\bibitem{Wahba1980}
Wahba, G.: Automatic smoothing of the log periodogram.
\newblock Journal of the American Statistical Association \textbf{75}(369),
  122--132 (1980).
\newblock \doi{10.1080/01621459.1980.10477441}

\bibitem{Wahba1990}
Wahba, G.: Spline Models for Observational Data.
\newblock Society for Industrial and Applied Mathematics (1990).
\newblock \doi{10.1137/1.9781611970128}

\bibitem{Whittle1954}
Whittle, P.: On stationary processes in the plane.
\newblock Biometrika \textbf{41}(3-4), 434--449 (1954).
\newblock \doi{10.1093/biomet/41.3-4.434}

\bibitem{Whittle1957}
Whittle, P.: Curve and periodogram smoothing.
\newblock Journal of the Royal Statistical Society: Series B (Methodological)
  \textbf{19}(1), 38--47 (1957).
\newblock \doi{10.1111/j.2517-6161.1957.tb00242.x}

\bibitem{Wilson2013}
Wilson, A.G., Adams, R.P.: Gaussian process kernels for pattern discovery and
  extrapolation.
\newblock In: Proceedings of the 30th International Conference on International
  Conference on Machine Learning, vol.~28, pp. 1067--1075 (2013)

\bibitem{Wood2011}
Wood, S., Rosen, O., Kohn, R.: Bayesian mixtures of autoregressive models.
\newblock Journal of Computational and Graphical Statistics \textbf{20}(1),
  174--195 (2011).
\newblock \doi{10.1198/jcgs.2010.09174}

\bibitem{Wood2002}
Wood, S.A., Jian, W., Tanner, M.: Bayesian mixture of splines for spatially
  adaptive nonparametric regression.
\newblock Biometrika \textbf{89}(3), 513--528 (2002).
\newblock \doi{10.1093/biomet/89.3.513}

\bibitem{Wood2017}
Wood, S.N.: P-splines with derivative based penalties and tensor product
  smoothing of unevenly distributed data.
\newblock Statistics and Computing \textbf{27}(4), 985--989 (2017).
\newblock \doi{10.1007/s11222-016-9666-x}

\end{thebibliography}
\end{document}